\documentclass[conference]{IEEEtran}
\IEEEoverridecommandlockouts
\usepackage{cite}
\usepackage{amsmath,amssymb,amsfonts}
\usepackage{algorithmic}
\usepackage{graphicx}
\usepackage{textcomp}
\usepackage{xcolor}
\def\BibTeX{{\rm B\kern-.05em{\sc i\kern-.025em b}\kern-.08em
    T\kern-.1667em\lower.7ex\hbox{E}\kern-.125emX}}
\begin{document}

\title{Contextual Beamforming: Exploiting Location and AI for Enhanced
Wireless Telecommunication Performance \\
}

\author{\IEEEauthorblockN{Jaspreet Kaur, Satyam Bhatti*, Olaoluwa R Popoola, Muhammad Ali Imran, Rami Ghannam, \\ Qammer H Abbasi and Hasan T Abbas}
\IEEEauthorblockA{\textit{James Watt School of Engineering, University of Glasgow, United Kingdom} \\
\textit{Email: \{j.kaur.1, s.bhatti.2*\}@research.gla.ac.uk}\\
\{Olaoluwa.Popoola, Muhammad.Imran, Rami.Ghannam, Qammer.Abbasi and Hasan.Abbas\}@glasgow.ac.uk}
}

\maketitle

\begin{abstract}
The pervasive nature of wireless telecommunication has made it the foundation for mainstream technologies like automation, smart vehicles, virtual reality, and unmanned aerial vehicles. As these technologies experience widespread adoption in our daily lives, ensuring the reliable performance of cellular networks in mobile scenarios has become a paramount challenge. Beamforming, an integral component of modern mobile networks, enables spatial selectivity and improves network quality. However, many beamforming techniques are iterative, introducing unwanted latency to the system. In recent times, there has been a growing interest in leveraging mobile users' location information to expedite beamforming processes. This paper explores the concept of contextual beamforming, discussing its advantages, disadvantages and implications. Notably, the study presents an impressive 53\% improvement in signal-to-noise ratio (SNR) by implementing the adaptive beamforming (MRT) algorithm compared to scenarios without beamforming. It further elucidates how MRT contributes to contextual beamforming. The importance of localization in implementing contextual beamforming is also examined. Additionally, the paper delves into the use of artificial intelligence schemes, including machine learning and deep learning, in implementing contextual beamforming techniques that leverage user location information. Based on the comprehensive review, the results suggest that the combination of MRT and Zero forcing (ZF) techniques, alongside deep neural networks (DNN) employing Bayesian Optimization (BO), represents the most promising approach for contextual beamforming. Furthermore, the study discusses the future potential of programmable switches, such as Tofino, in enabling location-aware beamforming.
\end{abstract}


\section{Introduction}
{E}{very} subsequent generation of cellular communication has brought advancements in data speeds and capabilities, with each generation offering significant improvements over its predecessor \cite{chataut2020massive}.  The first-generation (1G) introduced the concept of cell phones, while the second-generation (2G) enabled text messaging services. The advent of the third-generation (3G) brought about internet streaming capabilities, and the fourth-generation (4G) revolutionised the mobile landscape with broadband internet coverage. However, as user demands continue to escalate rapidly, 4G networks have reached their capacity limits, necessitating the need for more data to cater to the growing number of smartphones and smart devices. 

The arrival of fifth-generation (5G) cellular technology promises to address these challenges by providing networks capable of carrying significantly higher traffic volumes than currently available networks\cite{kumar2021review}. 5G networks are expected to evolve at a rate ten times faster than the long-term development of 4G (LTE), these have the potential to facilitate the development of technologies such as augmented reality (AR), autonomous vehicles, the internet of things (IoT) \cite{gohar2021role}. At the core of 5F technology are five key advancements: full-duplex, massive multi-input multi-output (MIMO), millimetre waves (mmWaves), smart cell, and beamforming (BF). While smartphones and electronic devices operate within radio frequency (RF) frequencies that are typically less than 6 GHz \cite{huo20175g, zhang2015large, alhayani20225g}. This range is becoming increasingly congested due to the proliferation of communication technologies and multiple mobile carriers. The limited RF spectrum available in the industrial, scientific and medical (ISM) band poses challenges for accommodating the growing demand for data transmission, resulting in slower services and more frequent lost connections\cite{sandoval2016evaluating,sethi2017internet}. 

To address this issue, researchers have explored higher frequency bands ranging from 30 to 300 GHz \cite{sandoval2016evaluating,sethi2017internet}. Although millimetre waves(mm-wave) have been utilized in satellite communication for some time, their use in mobile communications is a recent development. While offering a wider frequency spectrum, mm-waves face a major challenge due to their limited ability to penetrate obstacles such as infrastructure. This characteristic leads to signal loss or absorption when mm-waves encounter environmental obstacles \cite{rappaport2013millimeter}. Subsequently, smart cell networks provide a solution to overcome this problem. Unlike traditional cellular connections relying on large high-power cell towers that transmit signals over long distances, smart cell networks leverage thousands of small low-power access points (APs) \cite{attaran2021impact, pahlavan2001principles}. These APs are strategically placed in close proximity and grouped spatially to relay signals around obstructions. By eliminating the reliance on line of sight (LOS), smart cell networks ensure uninterrupted cellular service, even when users move behind obstacles. When a user equipment (UE) travels behind an obstruction, it seamlessly switches to a new AP, maintaining a consistent connection \cite{akyildiz2018combating, tripathi2021millimeter}. 

\begin{figure*}[!t]
\centering
\includegraphics[width=12cm]{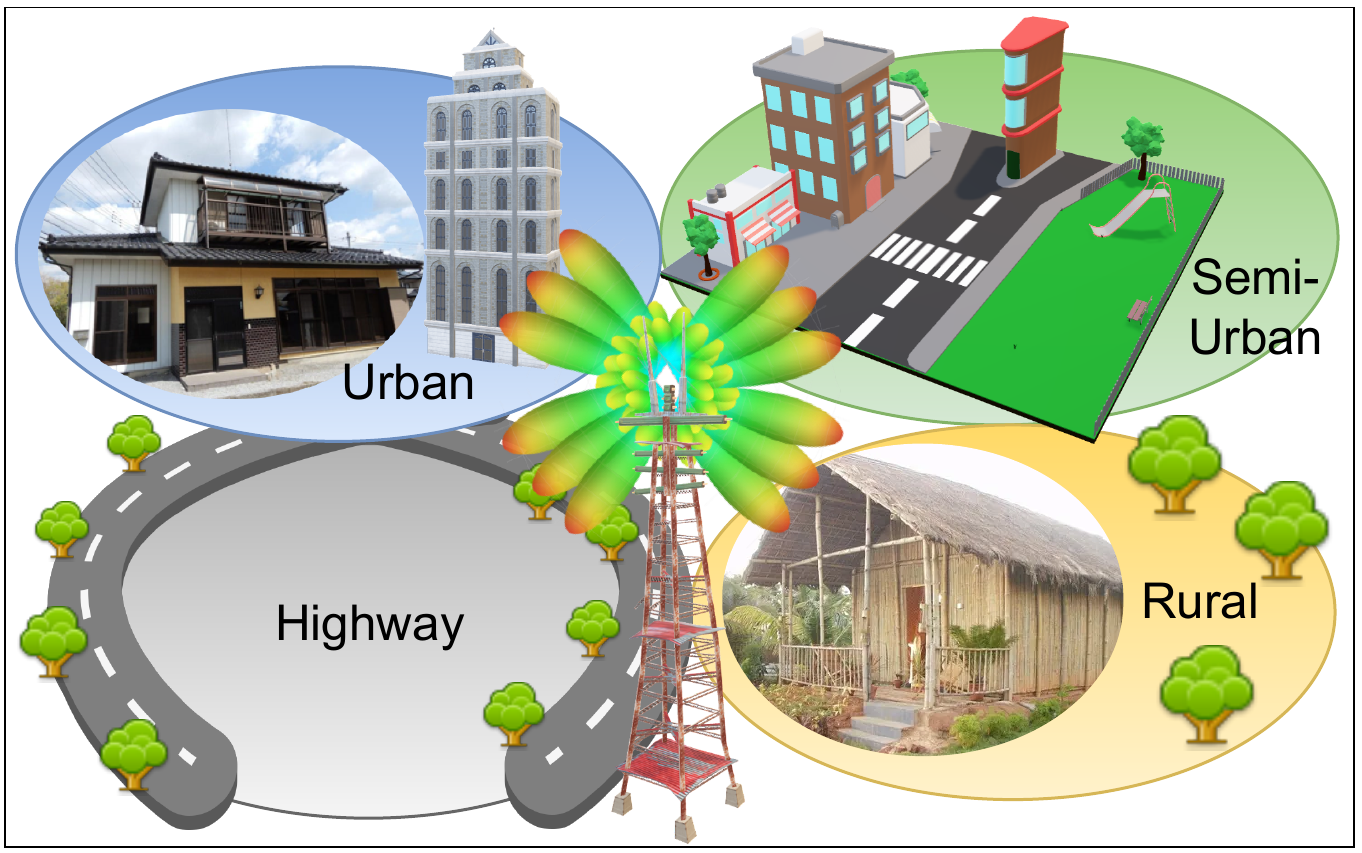}
\caption{Illustration of BF according to the scenario (rural, semi-urban, urban, highway).}
\label{scenario Specific BF}
\end{figure*}

Another significant advancement in 5G technology is the use of Massive MIMO, which involves deploying a higher number of antennas compared to traditional MIMO systems. Massive MIMO leverages beamforming techniques to direct wireless signals towards their intended receivers and enables spatial multiplexing of multiple data streams over the same frequency band. Despite its drawbacks, Massive MIMO can significantly enhance communication performance and multiply the capacity of a mobile ad-hoc network by a factor of 22 or more\cite{chataut2020massive,marzetta2010noncooperative}. In a time-division multiplexing system, user equipment (UE) needs to alternate between transmitting and receiving, which can introduce delays and reduce communication efficiency. In traditional cellular base stations (BS), antennas can only broadcast or receive signals at a given time. Multiplexing can improve performance, but transmit and receive signals are typically propagated at different frequencies. Conventional cellular antennas broadcast signals in all directions simultaneously, leading to potential interference \cite{zakrzewska2014towards}. Figure \ref{scenario Specific BF} provides an illustration of the beamforming process (broadcasting signals in a specific direction) in rural, semi-urban, urban, and highway areas.

Beamforming, on the other hand, offers several advantages in cellular communication. It enables more reliable and faster data transmission by establishing a more direct connection between transmitters and receivers. Beamforming has become an essential technology in various applications, including the 5G standard for cellular networks and radar-detection systems. However, implementing beamforming requires significant processing resources, which can pose challenges related to cost, hardware, and energy consumption. In the past, radar systems relied on mechanically moving and steering antennas to direct signals. However, advancements in signal processing techniques have made it possible to manipulate radio waves and focus them towards specific locations using electromagnetic beams. This eliminates the need for physical movements and reduces dependence on the physical structure of antennas.

The development of antenna systems for 5G networks must meet the requirements of compact size and low power consumption. To enhance spectrum efficiency and throughput, antenna arrays with larger dimensions, such as 64 x 64 MIMO and beyond, are being utilized. However, the accuracy of these antenna arrays significantly affects the performance of beamforming. As wavelengths decrease, component sizes, including RF transceivers with features like ADCs, also decrease. Exploring new materials, such as 40 nM CMOS, is helping to reduce the size and power consumption of essential components in 5G networks. Traditional RF power amplifiers made with materials like GaAs and other III-V semiconductors are not power-efficient and do not integrate well with other capabilities. This is where advancements in 40 nM CMOS technology can play a role in further reducing the size and power consumption of these critical components. Moreover, as the number of beams created by individual gNBs increases, more advanced signal processing techniques, including neural network methods, are required. This pushes power budgets and space restrictions even further. Despite these challenges, beamforming holds a promising future in various application areas.

Contextual beamforming is a promising technique for enhancing the performance of 5G communication systems. It enables the use of millimeter wave frequencies and massive MIMO technologies to achieve high data rates and low latency. Contextual beamforming adapts beamforming parameters in real-time based on environmental conditions and user requirements. This is achieved through feedback from the network and user devices, as well as the utilization of machine learning algorithms to optimize the beamforming process. Contextual Beamforming has applications in mobile edge computing (MEC), where low-latency computing and networking services are provided to mobile users. By dynamically adjusting the beamforming parameters based on the location, movement, and traffic conditions of the users, the quality of the wireless links between the user devices and the MEC servers can be improved. In VR/AR, contextual beamforming can improve the quality of audio and video streams used in VR/AR applications by selectively enhancing relevant signals and suppressing irrelevant or distracting ones.

To optimize the beamforming process and make it more efficient, AI is used in 5G technology. AI algorithms analyze real-time data from the network to determine the optimal beamforming pattern based on various factors, such as the user's location and behaviour, and the surrounding environment. Additionally, AI can help with beamforming in scenarios involving multiple users, where each user requires a different beamforming pattern. By dynamically adjusting the beamforming patterns for each user, optimal signal strength can be ensured. In complex scenarios involving multiple users and dynamic environments, AI is a crucial tool for optimizing the beamforming process in 5G networks.

\subsection{Contribution to the Literature}

In our review paper, we analyzed the role of Beamforming in the field of mobile communication. We conducted a review of the applications of beamforming in the algorithms, antenna fabrication, and the discovery of new AI (artificial intelligence) approaches. Our article is an effort to provide a review of contextual beamforming. The following are the major contributions of this article:

\begin{enumerate}
  \item We shortlisted research articles related to beamforming techniques that can help in context-aware beamforming.
    \item We review the literature on beamforming types both standard and using ML and AI techniques.
    \item Various ML techniques facilitated the estimation of user location and beam management in the study.
    \item We investigated the techniques used for the optimization of beamforming with the help of ML.
    \item We highlighted the challenges associated with using ML techniques for contextual beamforming.    
\end{enumerate}

\subsection{State of the Art}
The field of contextual beamforming in 5G technology is in a constant state of evolution, with ongoing research and development efforts aimed at improving its efficiency and effectiveness. In recent years, a surge in the use of conventional, machine learning (ML), and artificial intelligence (AI) techniques for beamforming has been observed. Several recent advances have been made in this area. By considering location-unaware systems with benchmarking techniques \cite{sand2009position}, a location-aware system can be developed and make location estimation more accurate. Additionally, opportunistic beamforming is used as feedback to smart antennas using channel delay information\cite{cheng2012location}.

In \cite{ruan2016low}, the authors propose a recursive matrix shrinkage method to estimate the interference-plus-noise covariance matrix along with the desired signal steering vector mismatch. A two-stage design approach was utilized in \cite{bogale2020adaptive}, with the first stage dealing with beamforming, and the second with adaptive power allocation and modulation. Another recent study by \cite{maschietti2017location} proposed a novel and general approach to deriving the statistical distribution of the signal-to-noise ratio (SNR) by exploiting the array structure, beamforming type, and slow fading channel coefficients. This approach was used to design power and modulation adaptation strategies. \cite{igbafe2019location} presented the scheme  that uses coordinated beam search from a small beam dataset within the error offset, and then the selected beams are used to guide the search for beam prediction.

Additionally, \cite{myers2020deep} proposed an end-to-end deep learning technique to design a structured compressed sensing (CS) matrix that is well-suited to the underlying channel distribution. This technique leverages sparsity and the spatial structure that appears in vehicular channels. In contrast, \cite{li2018generative} noted that current millimetre-wave (mmWave) beam training and channel estimation techniques do not typically make use of prior beam training or channel estimation observations. Moreover, \cite{alkhateeb2018deep} presented that determining the optimal beamforming vectors in large antenna array mmWave systems necessitates significant training overhead, which can have a significant impact on the efficiency of these mobile systems. However, a significant drawback in these studies was that limited ML approaches were discussed, and the optimization scope and implementation of proposed beamforming algorithms were not considered in real environments.

As various ML techniques have been adopted for BF, this paper aims to provide a detailed review of different ML-based BF techniques. These ML techniques include the procedure to preprocess the input data and various ML algorithms in any environment. Our systematic review goes beyond existing literature, showcasing how various ML techniques can be used to screen large numbers of beamforming approaches for potential location estimation applications and to optimize the approaches using high computational power. Accordingly, the following sections will describe the in-depth analysis of currently popular beamforming techniques and how AI can improve their overall performance by mitigating their limitations.

\subsection{Organisation of the Review}

The rest of the paper is organized depending on finding the gap from the state-of-the-art in the recent literature review and is described as follows. The adopted methodology is presented in section 2 of the systematic review, followed by the results and analysis of the systematic review are discussed in section 3. In addition, section 4 elaborates on the challenges associated with using the recent AI, ML, and DL techniques for contextual beamforming. Moreover, section 5 incorporates the potential applications of contextual beamforming. Subsequently, section 6 discusses our key contributions towards the development of Localisation and Beamforming. Lastly, section 7 of the systematic review concludes the article.

\begin{figure*}[!t]
 \centering
 \includegraphics[width=17cm]{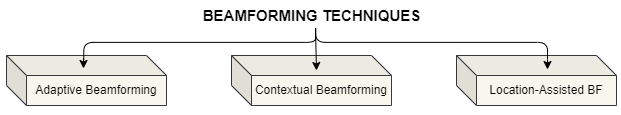}
 \caption{Base station to vehicle scenario.}
 \label{fig:block}
\end{figure*}

\section{Review Methodology}
This section of the systematic review presents our review methodology depending on the defined research objectives and questions that were used for shortlisting the relevant research articles on ML algorithms for contextual beamforming techniques.  

\subsection{Research Objectives}
The four key objectives of our systematic review article are:

\begin{enumerate}
    \item [RO1:] To review the range of Adaptive beamforming and ML-based beamforming using priori user data.
    \item [RO2:] To identify the ML techniques used specifically for Contextual beamforming.
    \item [RO3:] From a practical perspective, identify the specific ML and optimization techniques used for real-time implementation. 
    \item [RO4:] To identify ML algorithms specifically used for the beamforming for low latency, high throughput and SINR. 
\end{enumerate}

\subsection{Research Questions}
Our systematic review aims to answer the following four research questions:

\begin{enumerate}
  \item [RQ1:] What are the various location-assisted BF techniques?
  \item [RQ2:] What are the datasets required for classifying the Contextual Beamforming?
  \item [RQ3:] How can the Contextual Beamforming models be optimized for real-time processing?
  \item [RQ4:] What are the different types of ML and AI techniques used with respect to Contextual Beamforming?
\end{enumerate}

\subsection{Review Protocol}
For structuring our systematic review, we instigated a review protocol, and the following are the perquisites of the adopted analogy. In this section, we discuss the search strategy, inclusion criteria, exclusion criteria, and screening mechanisms for selecting relevant research papers.

\subsubsection{Search Strategy}
The most recent research papers from renowned publishing houses like IET, Science Direct, Nature, AIP, Wiley, IEEE Explorer, IoP science, ACS publications, and MDPI were taken into account during our review. We also included unreviewed papers from arXiv in our search. As a result, we evaluated and critically analysed the grey literature (research and publications produced by organisations not usually linked with academic or commercial publishing organisations) using the AACODS (Authority, Accuracy, Coverage, Objectivity, Date, Significance) criteria. 

We commence by queering every repository that contains various study items. To compile our study articles, we defined keywords like "Machine Learning," "Deep Learning," "Beamforming," "location," "context information," "5G," and "Vehicular Communication." Based on the article's title and abstract as well as a full-text read of the papers, articles were scanned. To link these keywords, we also created search strings using the Boolean operators AND and OR. 
\subsubsection{Inclusion Criteria}
The following are the parameters used in the inclusion criteria.

\begin{enumerate}
    \item We included only English-language articles involving the data-driven approaches of beamforming using conventional and ML techniques and were pertinent to the study issues such as poor data quantity and data quality.
    \item We included the pertinent articles facilitating the discovery of only low-latency beamforming algorithms using ML methods before determining their eligibility.
   \item We included comparative studies involving the optimization and robustness of beamforming techniques designed from ML services.
    \item We targeted only articles that discussed ML for beamforming, location information, and publications on ML integration on contextual beamforming.
\end{enumerate}

\subsubsection{Exclusion Criteria}

The following is a list of the exclusion criteria for shortlisting the research papers based on our research objectives and targeted research questions.

\begin{enumerate}
    \item English-language research articles released in other languages. 
    \item Research papers without a complete text version. 
    \item Editorials, review articles of surveys, abstracts, and short papers concerning secondary studies are not accepted.
    \item Articles that didn't discuss how to combine ML techniques with beamforming.
\end{enumerate}

\subsubsection{Screening Phase}

Articles were further screened in two phases. In the first phase, we examined the title and the abstract of each research article to check whether they satisfied our inclusion criteria. In the second phase, we further shortlisted our articles based on their full text. It is worth mentioning that the same piece of writing frequently appeared in various publications. For example, conference papers frequently appear in journals. We take into account the original writing each item was reviewed throughout the screening stage two. At least two of the contributors of this paper who were entrusted with classifying the items as either pertinent or not pertinent might require more research, as finalized until any such item is either published or the authors have a discussion tagged as relevant or not. Survey and review papers were excluded from our review. Finally, each article was carefully classified and evaluated thematically. 

\section{Results and Analysis of the Review}

This section of the review paper summarizes the research articles that are shortlisted using the defined research objectives as well as aims to answer the predefined research questions. 

\subsection{What are the recent BF techniques [RQ:1]}

With an extensive utilization of GPS coordinates worldwide, various location-assisted BF techniques are becoming exponentially important. Contextual beamforming and adaptive beamforming are two techniques used in signal processing to improve the quality of transmitted or received signals. Contextual beamforming refers to using prior knowledge about the environment to design a beamforming algorithm that optimizes the signal quality in that specific environment. This prior knowledge can include information about the location and number of signal sources, the acoustic properties of the environment, and other factors that can affect the quality of the received signal. On the other hand, adaptive beamforming refers to using real-time feedback from the received signal to continuously adjust the beamforming algorithm to improve the quality of the signal. This is particularly useful in dynamic environments where the signal sources or environmental conditions may change over time. Contextual beamforming is based on prior knowledge of the environment, while adaptive beamforming uses real-time feedback to continuously adjust the beamforming algorithm. Both techniques have their strengths and weaknesses.

\begin{figure}[!t]
    \centering
    \includegraphics[width=\columnwidth]{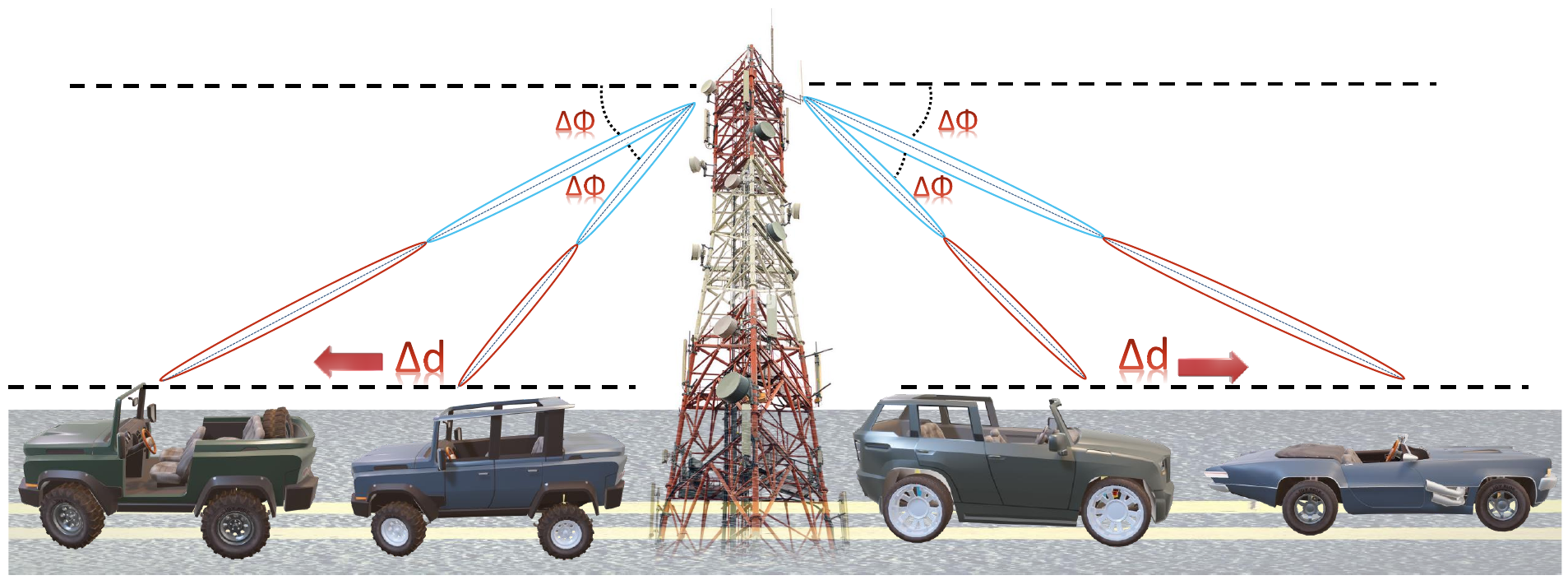}
    \caption{Base station to vehicle scenario}
    \label{car 2 sides}
\end{figure}

\subsubsection{Adaptive Beamforming}

An adaptive beamformer is a tool for performing adaptive spatial signal processing using an array of transmitters or receivers. The signals are integrated in such a way that the signal intensity to and from a specific direction is increased.  Signals from and to other directions are combined constructively or destructively, resulting in the degradation of the signal from and to the undesired direction. This method is utilised in both RF and acoustic arrays to achieve directional sensitivity without physically changing the receivers or transmitters \cite{blogh2002third, monzingo2004introduction, li2010mimo}. 

Adaptive BF was first developed in the 1960s for military sonar and radar applications. There are various modern applications for BF, with commercial wireless networks such as long-term evolution (LTE) being one of the most popular. Adaptive BF's first applications in the military were primarily focused on radar and electronic countermeasures to counteract the effects of signal jamming. In phased array radars, BF can be seen. These radar applications use either static or dynamic/scanning BF; however, they are not truly adaptive. Adaptive BF is used in commercial wireless standards such as 3GPP LTE and IEEE 802.16 WiMAX to enable important services within each standard \cite{steinhardt1989adaptive}. The concepts of wave transmission and phase relations are used in an adaptive BF system. A greater or lower amplitude wave is formed, for example, by delaying and balancing the received signal, using the concepts of superimposing waves. 

\begin{figure}[!t]
    \centering
    \includegraphics[width=\columnwidth]{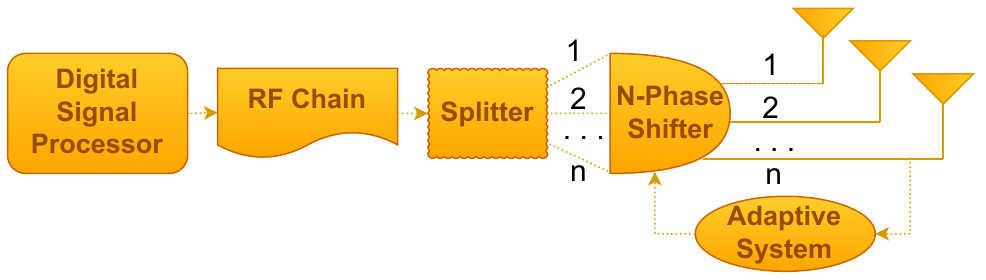}
    \caption{Basic block diagram of Adaptive BF}
    \label{ADAPBF}
\end{figure}

The adaptive BF system is adaptive in real-time to maximize or minimize desirable parameters, including the signal-to-interference noise ratio (SINR). There are numerous approaches to BF design, the first of which was achieved by Applebaum in 1965 by increasing the signal-to-noise ratio (SNR) \cite{shaikh2015linear}. This method adjusts the system parameters to maximize the power of the received signal while reducing noise (jamming or interference). Widrow's least mean squares (LMS) error method and Capon's maximum likelihood method (MLM) introduced in 1969 are two further approaches. The Applebaum and Widrow algorithms are quite similar in that they both converge on the best option. However, these strategies have difficulties in terms of implementation. Reed demonstrated a technique called sample matrix inversion (SMI) in 1974 \cite{reed1974rapid}. Unlike Applebaum and Widrow's approach, SMI determines the adaptive antenna weights directly \cite{blogh2002third,monzingo2004introduction,li2010mimo}.

The Weiner solution can be used to create statistically optimal weight vectors for adaptive BF in data-independent BF design methods. On the other hand, the asymptotic $2^{nd}$ order statistics of SINR were assumed. Statistics fluctuate over time in cellular networks where the target is mobile and interferes with the cell area. An iterative update of weights is required to follow a mobile user in a time-varying signal propagation environment \cite{vorobyov2014adaptive}. This enables the spatial filtering beam to adjust to the time-varying DOA of the target mobile user and to provide the desired signal to the user. To address the challenge of statistics (which can vary over time), adaptive algorithms that adapt to changing environments are frequently used to determine weight vectors. The functional block diagram of an adaptive array of n elements includes an antenna array of n elements and a digital signal processor with a feedback and/or control loop algorithm. The signal processing unit receives the data stream gathered by an array and computes the weight vector using a specific control method.

On the contrary, the adaptive antenna array is divided into two categories: a) steady-state and b) transient state. These two categories are determined according to the array weights of stationary environment and time-varying environment. If the reference signal for the adaptive method is known from prior information, the system can update the weights adaptively through feedback \cite{lukose2010study}. To change the weights of the time-varying environment at every instance, several adaptive algorithms (mentioned in the further section) can be utilized. Figure \ref{ADAPBF} shows the block diagram for adaptive BF which consist of a digital signal processor (DSP), RF chain, splitter, and N-phase shifter followed by antenna assembly along with an adaptive system providing feedback to shifters.

\begin{figure}[!t]
    \centering
    \includegraphics[width=\columnwidth]{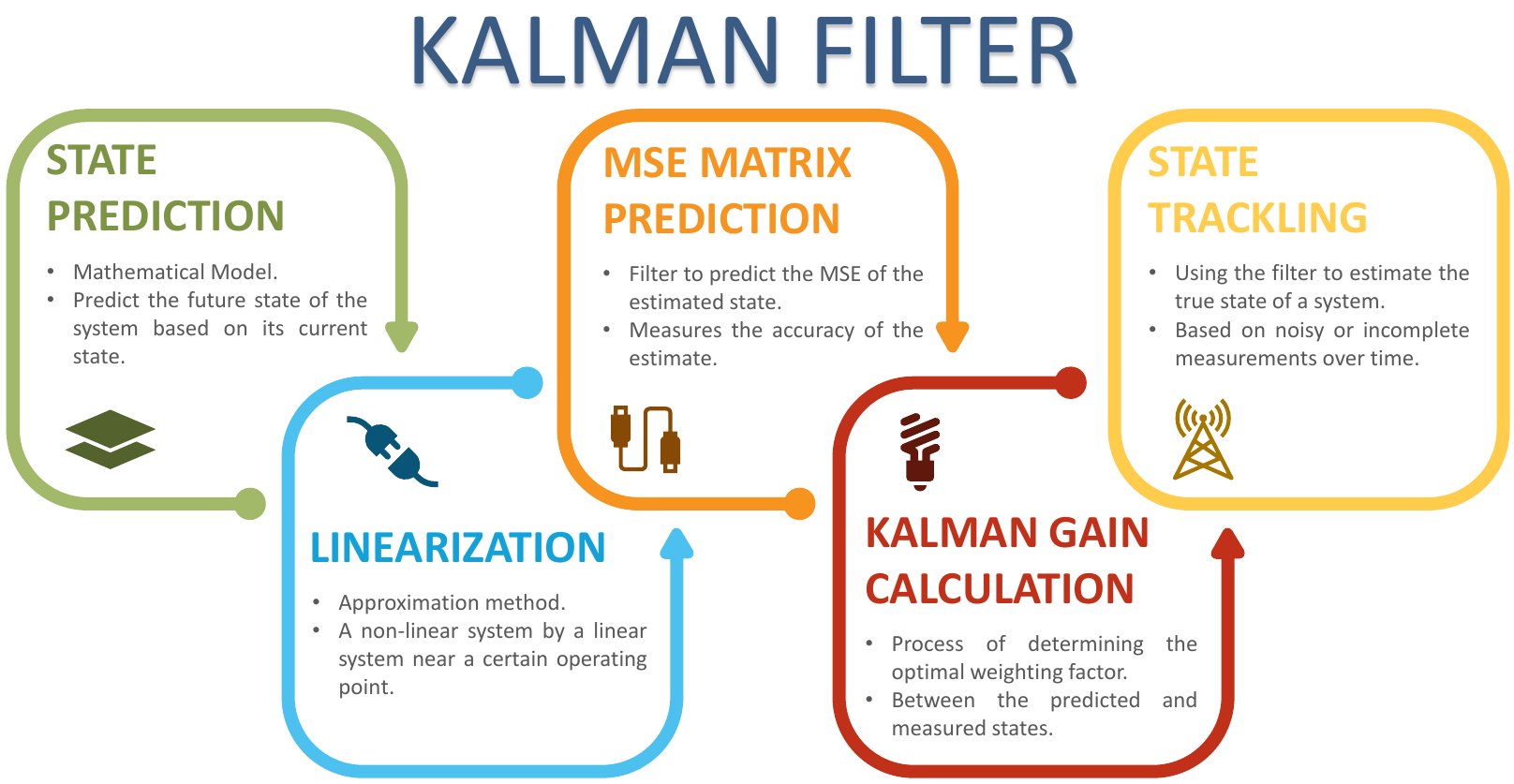}
    \caption{A standard procedure based on the Kalman filter.}
    \label{kalman filter}
\end{figure}

\subsubsection{Contextual Beamforming}
The ability to forecast the next location of the receiver which is based on tracking previous movements can be useful for creating intelligent applications like automobiles, robotics, augmented/virtual reality etc. The advancement of location prediction apps and services is enabled by the growth of methodologies for predicting and projecting the receiver's position in the future \cite{werner2002reduced}. A wireless system, in general, controls a location-predicting framework by capturing and communicating critical data before application. The sender must be able to determine the receiver's location at any given time to interact effectively with them. Machine learning (ML) methods have already been used to predict the receiver's location. Context is created by recording, processing, and transcribing the receiver's status data at a certain time. Several machine learning algorithms, such as Deep Neural Networks, Convolution Neural Networks, Generative Adversarial Networks, and others, have been recognised as aiding in the technological advancement of location forecasting. Furthermore, depending on the application, machine learning algorithms can be modified and customised to match their objectives \cite{orikumhi2018location}.

The majority of the existing mmWave beam tracking research focuses on communication-only protocols. The unusual beam tracking technique requires the transmitter to send information to the receiver, which then determines the angular position and sends it back to the transmitter. It is worth noting that in high-mobility communication circumstances, such as the one depicted in Figure \ref{car 2 sides}, it is not enough to merely track the beam. To meet the crucial latency requirement, the transmitter should be capable of predicting the beam \cite{dai2021adaptive}. The state prediction and tracking designs in Figure \ref{kalman filter} is based on the classic Kalman filtering process.

\begin{figure}[!t]
    \centering
    \includegraphics[width=\columnwidth]{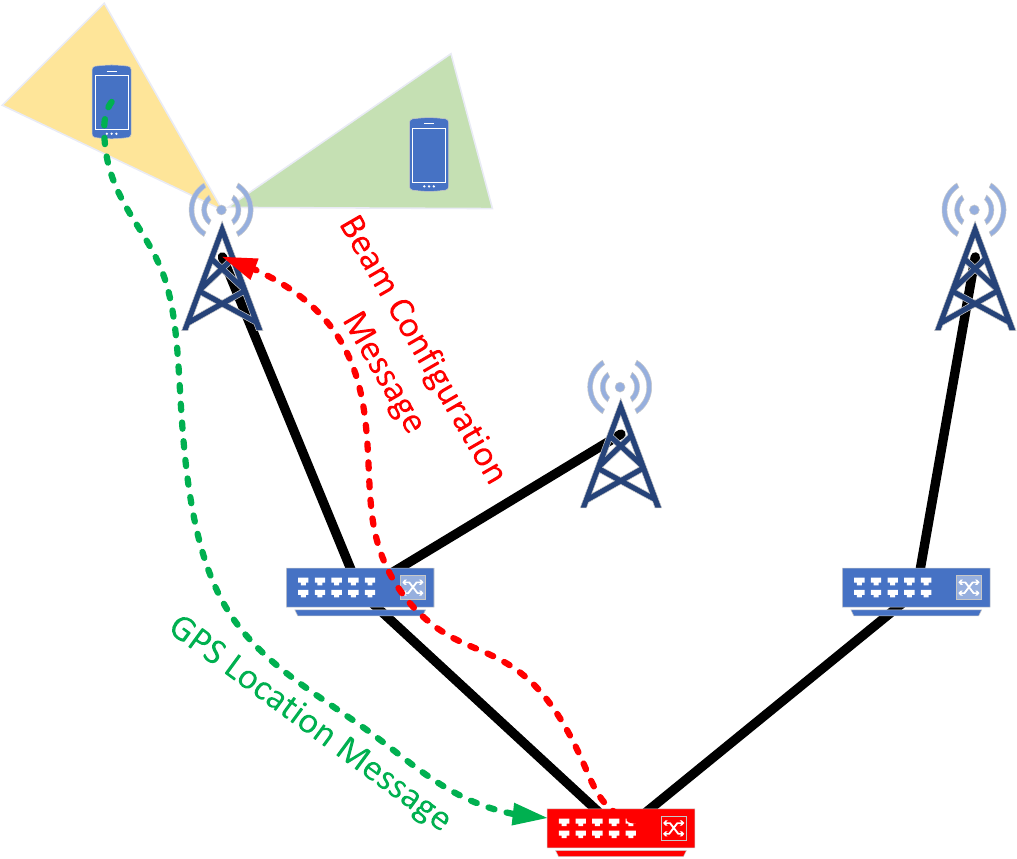}
    \caption{Basic block diagram of Location-Assisted BF}
    \label{crop}
\end{figure}

\subsubsection{Location-assisted Beamforming}
\begin{figure}[!t]
    \centering
    \includegraphics[width=\columnwidth]{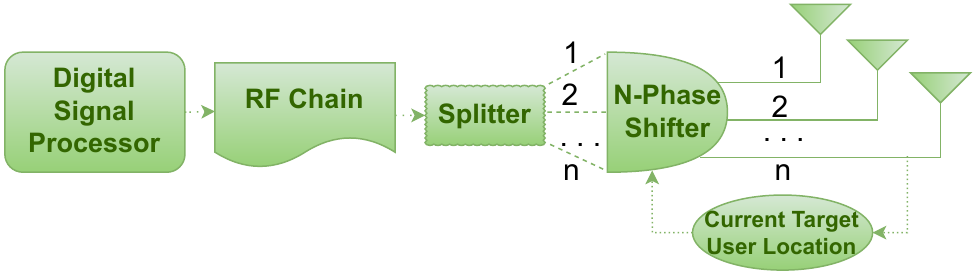}
    \caption{Basic block diagram of Location-Assisted BF}
    \label{LBF}
\end{figure}

The a priori information on the location of the user can enable the system to work more efficiently. The sorting of the prior information can reduce energy footprints. As an example, the branch predictor \cite{parihar2015branch} in computer architectures can improve the flow in the instruction pipeline to achieve highly effective performance. In the case of location-aided or location-aware BF, a similar concept has been seen. Figure \ref{LBF} shows the block diagram for predictive or location-assisted BF which consist of a digital signal processor (DSP), RF chain, splitter, and N-phase shifter followed by antenna assembly along with a feedback loop providing current target user location to shifters. Line of sight (LOS) communication in mmWave transmission systems provides multi-gigabit data transmission with BF toward the user direction to mitigate the substantial propagation loss. However, abrupt performance degradation caused by human obstruction remains a major issue, thus using possible reflected pathways when blocking occurs should be considered \cite{kim2018development}.

In this line, location is a critical factor for Contextual beamforming in 5G networks because 5G relies on higher frequencies and smaller cells than previous generations of wireless networks. These higher frequencies have shorter wavelengths, which means that the signal is more easily obstructed by obstacles such as buildings, trees, and other objects. As a result, the location of the user and the position of the cell tower are crucial for ensuring reliable and efficient communication.
Contextual beamforming in 5G networks involves directing the transmission and reception of signals towards the user's location, which can significantly improve signal quality and reduce interference from other sources. This can be done using beamforming techniques, which use phased array antennas to focus the signal in a specific direction towards the user. By adjusting the direction and shape of the beam, beamforming can improve signal strength and quality, reduce interference, and increase the capacity of the network. In addition to beamforming, 5G networks also rely on other location-based technologies, such as geolocation and network slicing. Geolocation can be used to determine the user's location and provide location-specific services, while network slicing allows for the creation of virtualized networks with different performance characteristics for different locations and applications.

The usage of location-aware beamforming (BF) and interference mitigation techniques in ultra-dense 5G networks composed of densely scattered access nodes (AN) has been investigated in the literature. The development of user environment area networks (UEAN) with short distances in a packed environment results in higher levels of signal interference, but network densification enhances the chance of line-of-sight (LoS) and, as a result, leads to more accurate UE placement. This enables the use of spatial dimensions by BF and interference reduction. The accuracy of radio network positioning systems currently available is inferior to that of fibre optic communication systems in radar stations and atomic clock-based satellite navigation systems. Future 5G networks are expected to provide positioning accuracy on the order of one meter. Lu et al \cite{sand2009position} proposed approaches such as weighted centroid geometric (WCG) and a joint positioning and tracking framework based on the extended Kalman filter (EKF) to achieve accurate and reliable 3D positioning for industrial IoT systems where anchor locations are not precisely known. They also suggested position-aided beamforming (PA-BF) approach that outperforms conventional BF in terms of initial access latency and spectral efficiency, especially for UE moving at a speed greater than 0.6 m/s.

Sellami et al \cite{sellami2021outdoor} proposed a neighbour-aided localization algorithm for outdoor UEs operating in challenging channel conditions. The algorithm selects two neighbours based on reference signal power measurement, and the BS performs beamforming over an angular interval determined by the calculated distance and angle of arrival (AoA) of the first neighbour to discover two candidates for the UE post. \cite{wang2021location} provided location assistance (LA) direction of departure (DOD)-based beamforming technique that is appropriate for wireless communication in high-speed rail (HSR). The algorithm's goal is to modify the phase at the transmitter to increase the output signal-to-noise ratio (SNR) at the receiver. Both the performance of ideal DOD beamforming and approximated DOD with location error-related variation are assessed.

The study described in \cite{dobler2019lama} suggests LAMA (Location-Assisted Medium Access), a Medium Access Control (MAC) protocol based on locally shared position data for position awareness beaconing. Their contention-free method manages to effectively minimize interference, especially hidden-terminal type, through coordinated spatial reuse and scales effectively with high neighbour numbers. In \cite{lazarev2019positioning}, the authors implemented location-aware beamforming and interference mitigation techniques in 5G ultra-dense radio networks to improve the use of space. They also estimated the positioning accuracy limitations of the user equipment using the direction of arrival measurement processing in three-dimensional space with metrics of the Cramer-Rao lower bound ellipsoid.

Similarly, \cite{xing2021location} proposed a location-aware beamforming design for the RIS-aided millimetre-wave (mmWave) communication system without the channel estimation process, which took into account the limitations of the conventional channel state information (CSI) acquisition techniques for the RIS-aided communication system. They also created a worst-case robust beamforming optimization problem to counteract the impact of location inaccuracy on the beamforming design. In \cite{mohammadi2022location}, a spatial estimation approach based on the theory of observers in the control systems literature was proposed to handle the essential issue of beamforming in UAV-based communications. They effectively anticipated the positions of the target UAVs in the presence of uncertainty by using a delay-tolerant observer-based predictor. The method functioned consistently in the presence of channel blockage and interference.

The likelihood of positioning-aided beamforming systems experiencing an outage was investigated in \cite{zhu2022outage}. The authors took into consideration positioning error, link distance, and beamwidth to generate closed-form outage probability constraints. They demonstrated that the beamwidth should be maximized with the transmit power and connection distance to reduce the likelihood of an outage. In \cite{liu2020location}, a deep learning-based location-aware predictive beamforming technique was proposed to follow the beam for UAV communications in a dynamic environment. They developed a long short-term memory (LSTM)-based recurrent neural network (LRNet) to predict the UAV's expected location, which could be used to calculate a forecast angle between the UAV and the base station for efficient and quick beam alignment.

In a multi-cell, multiple input, multiple outputs (MIMO) communication system aided by optical positioning, \cite{wang2021location} suggested a location-based energy-efficient optimization approach for the beamforming matrix. They increased the system's achievable ergodic rate by estimating the channel coefficient matrix based on the location data. In \cite{lu2020positioning}, position-aided beamforming (PABF) architecture was proposed for improved downlink communications in a cloud-oriented mmWave mobile network. The authors demonstrated that the proposed PABF outperformed the traditional codebook-based beamforming in terms of effective transmit ratio and initial access latency, demonstrating its potential to accommodate high-velocity mobile users.

Finally, \cite{orikumhi2018location} proposed an effective beam alignment solution for mmWave band communications by utilizing the mobile user's location data and potential reflectors. The suggested method enabled the base station and mobile user to jointly search a small number of beams within the error bounds of the noisy location information. Additionally, \cite{khosravi2022location} proposed a method for beamforming that tracked the spatial correlation of the strong pathways that were currently accessible between the transmitter and the receiver. They demonstrated the robustness of their approach to position information uncertainty and how it could reliably maintain a connection with a user who was travelling along a trajectory.

In summary, all the above-mentioned research shows the effectiveness of knowing the user location either static or on the move to leverage the existing cellular communications. The location information can assist in predicting the precoding weights accurately and eventually steering the beam in the direction of the user with low latency and high throughput.

\begin{figure*}[!t]
 \centering
 \includegraphics[width=12cm]{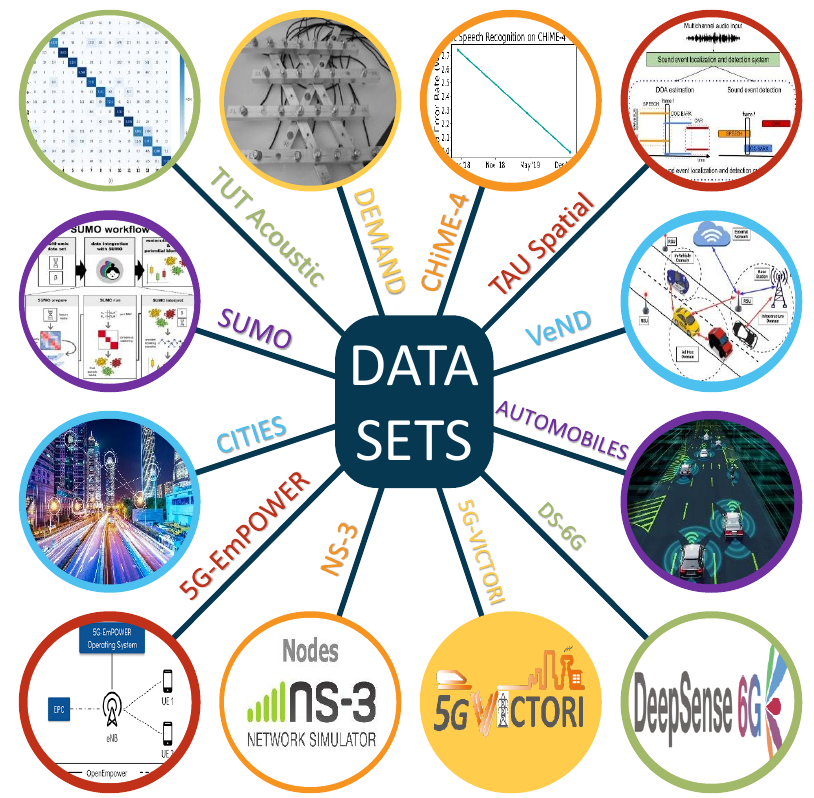}
 \caption{Datasets for Contextual Beamforming Classification}
 \label{fig:Ref1}
\end{figure*}

\subsection{Datasets for Contextual Beamforming Classification [RQ:2]}

Researchers often need datasets that contain location data, audio signals, and other pertinent elements to identify contextual beamforming approaches. The following are a few examples of datasets that have been applied in earlier research:

\subsubsection{\textbf{TUT Acoustic Scenes 2017 dataset}:} This dataset includes location-specific metadata and audio recordings of diverse acoustic scenes, including street traffic, parks, and retail malls. It has been employed in numerous research to gauge how well contextual beamforming methods function \cite{mesaros2016tut}.\cite{mesaros2018multi} introduced the acoustic scene classification task of DCASE 2018 Challenge and the TUT Urban Acoustic Scenes 2018 dataset provided for the task, and evaluates the performance of a baseline system in the task. The TUT Urban Acoustic Scenes 2018 dataset consisted of 10 different acoustic scenes recorded in 6 large European cities, making it more acoustically variable than previous datasets used for this task. The dataset included high-quality binaural recordings as well as data recorded with mobile devices. The baseline system consisting of a convolutional neural network achieved good performance in the subtasks using the recommended cross-validation setup. Also, \cite{mesaros2016tut} introduced the TUT Acoustic Scenes 2016 database, which was a collection of binaural recordings from 15 different acoustic environments. A subset of this database called the TUT Sound Events 2016, was annotated to mark sound events. The paper presented the recording and annotation procedure, the database content, and the performance of a supervised acoustic scene classification system and event detection baseline system.
     
\subsubsection{\textbf{DEMAND dataset}:} The DEMAND dataset includes audio recordings of metropolitan settings, including as parking lots, train stations, and street traffic. Associated metadata is also included, such as GPS positions and noise levels. Studies to assess the efficacy of contextual beamforming methods have utilised this dataset \cite{thiemann2013diverse}.    \cite{thiemann2013diverse} introduced the Diverse Environments Multi-channel Acoustic Noise Database (DEMAND), which was a set of 16-channel noise files recorded in a variety of indoor and outdoor settings. The data was recorded using a planar microphone array consisting of four staggered rows, with the smallest distance between microphones being 5 cm and the largest being 21.8 cm.
    
\subsubsection{\textbf{The CHiME-4 dataset}:} This collection of audio recordings comprises conversations that were overheard in places like offices, homes, and cafes. Additionally, it contains metadata like noise levels and position data. Studies evaluating the effectiveness of contextual beamforming methods for enhancing speech recognition accuracy have made use of the CHiME-4 dataset \cite{vincent20164th}.  This dataset is used in some research such as in \cite{schrank2016deep}, a deep eigenvector beamformer was proposed as a front-end for robust speech recognition in adverse environments. Data augmentation was performed by modulating the amplitude and time scale of the audio. The proposed system achieved a word error rate of 4.22\% on the real development and 8.98\% on the real evaluation data for 6 channels and 6.45\% and 13.69\% for 2 channels, respectively. Also, in \cite{christensen2010chime}, the CHiME corpus was designed to enable noise-robust speech processing research. The corpus included 40 hours of background recordings from a head and torso simulator in a domestic setting and a comprehensive set of binaural impulse responses. The data had been mixed to produce a controlled and natural range of SNRs for speech separation, enhancement, and recognition algorithms

\subsubsection{\textbf{Dataset for TAU Spatial Sound Events 2019}:} This collection includes location-related metadata and audio recordings of a variety of spatial sound occurrences, including footfall, glass breaking, and doors closing. Studies have used it to gauge how well contextual beamforming systems perform at locating and detecting sound occurrences \cite{adavanne2019multi}.     In \cite{politis2022starss22}, The STARS22 dataset was a high-resolution dataset of spatial recordings of real scenes with sound event annotations. The dataset was captured with a high-resolution spherical microphone array and delivered in two 4-channel formats, ﬁrst-order Ambisonics and tetrahedral microphone array.     Also, \cite{mesaros2016tut} introduced the TUT Acoustic Scenes 2016 database, which was a collection of binaural recordings from 15 different acoustic environments. A subset of this database called the TUT Sound Events 2016, was annotated to mark sound events. The paper presents the recording and annotation procedure, the database content, and the performance of a supervised acoustic scene classification system and event detection baseline system.

\subsubsection{\textbf{Vehicular Networks Dataset (VeND)}:} The University of California, Los Angeles (UCLA) created this dataset, which includes observations from a vehicular network testbed. The dataset contains data about the cars and their movements in addition to details about the wireless channel, such as the signal-to-noise ratio (SNR) and the channel impulse response (CIR) \cite{khelifi2019named}. 
    
   \cite{uppoor2013generation} presented a realistic synthetic dataset, covering 24 hours of car traffic in a $400-{km}^2$ region around the city of koln, in Germany. The dataset captures both the macroscopic and microscopic dynamics of road traffic over a large urban region. Incomplete representations of vehicular mobility may result in over-optimistic network connectivity and protocol performance.

\subsubsection{\textbf{5G-VICTORI}:} is a project financed by the European Union that aims to create 5G technologies for a range of applications, including vehicular communication. With regard to vehicular communication, the project has created a number of datasets, including assessments of the radio frequency (RF) channel and network performance in practical settings \cite{mesogiti20215g}. 
    
    \cite{bassbouss20215g}  discussed how the new 5G network technology would impact the digitalization of various industries, including modern railway transportation. The Future Railway Mobile Communication System (FRMCS) service requirements and system principles were well-mapped to 5G concepts, but deployment paradigms needed to be established to prove their effectiveness. The 5G-VICTORI project aimed to deliver a complete 5G solution for railway environments and FRMCS services, and this paper discussed the Key Performance Indicators and technical requirements for an experimental deployment in an operational railway environment in Greece.

\subsubsection{\textbf{5G-EmPOWER}:} This project, which is also supported by the European Union, aims to develop 5G technology for a range of applications, including vehicular communication. With regard to vehicular communication, the project has created a number of datasets, including assessments of the RF channel and network performance in practical settings \cite{coronado20195g}. 
    
    3GPP is embracing the concept of Control-User Plane Separation (a cornerstone concept in SDN) in the 5G core and the Radio Access Network (RAN). An open-source SDN platform for heterogeneous 5G RANs has been introduced, which builds on an open protocol that abstracts the technology-dependent aspects of the radio access elements. The effectiveness of the platform has been assessed through three reference use cases: active network slicing, mobility management, and load-balancing \cite{coronado20195g}.

\subsubsection{\textbf{NS-3}:} The Network Simulator 3 (NS-3) is an open-source network simulator that is useful for simulating and modelling vehicular communication in 5G networks. In addition to mobility models for simulating the movement of vehicles, NS-3 has various built-in modules for modelling the wireless channel \cite{riley2010ns}. 
    
    \cite{perrone2013design} presented a framework for the ns-3 network simulator for capturing data from inside an experiment, subjecting it to mathematical transformations, and ultimately marshalling it into various output formats. A framework for capturing data from inside an experiment, subjecting it to mathematical transformations, and ultimately marshalling it into various output formats is presented. The application of this functionality is illustrated and analyzed via a study of common use cases. The design presented provides lessons transferrable to other platforms.

\subsubsection{\textbf{Connected automobiles and Cities}:} The National Renewable Energy Laboratory (NREL) created this dataset, which contains information from a field investigation of connected automobiles in a smart city setting. The dataset contains details about, among other things, network performance, traffic flow, and vehicle trajectories\cite{sperling2019mobility}. 
    
    In \cite{tosi2017cell}, big data from the cellular network of the Vodafone Italy Telco operator can be used to compute mobility patterns for smart cities. Five innovative mobility patterns have been experimentally validated in a real industrial setting and for the Milan metropolitan city. These mobility patterns can be used by policymakers to improve mobility in a city, or by Navigation Systems and Journey Planners to provide final users with accurate travel plans.

\subsubsection{\textbf{DeepSense6G}:} DeepSense 6G is a collection of data that includes different types of sensing and communication information, such as wireless communication, GPS, images, LiDAR, and radar. This data was gathered in real-life wireless environments and represents the world's first large-scale dataset of this kind. The dataset contains over one million samples of this multi-modal sensing-communication data and was collected in over 30 different scenarios to target various applications. The collection of data was done at several indoor and outdoor locations with high diversity and during different times of the day and weather conditions. Additionally, there are tens of thousands of data samples that have been labelled both manually and automatically. 
     
     Also in \cite{alkhateeb2022deepsense},  the DeepSense 6G dataset is a large-scale dataset based on real-world measurements of co-existing multi-modal sensing and communication data. The DeepSense dataset structure, adopted testbeds, data collection and processing methodology, deployment scenarios, and example applications are detailed in the paper. The paper aims to facilitate the adoption and reproducibility of multi-modal sensing and communication datasets. The researchers (\cite{uvaydov2021deepsense}) had a 400 GB dataset containing hundreds of thousands of WiFi transmissions collected "in the wild" with different Signal-to-Noise Ratio (SNR) conditions and over different days. They also had a dataset of transmissions collected using their own software-defined radio testbed, and a synthetic dataset of LTE transmissions under controlled SNR conditions. 
     
\subsubsection{\textbf{SUMO}:} SUMO (Simulation of Urban Mobility) is an open-source traffic simulation software that allows modelling and simulating traffic flow in urban areas. It can simulate individual vehicles, pedestrians, public transportation, and various road networks. SUMO has a variety of applications, including traffic planning, intelligent transportation systems, and autonomous driving. A synthetic dataset generator was developed to support research activities in mobile wireless networks. The generator uses traces from the Simulation of Urban MObility (SUMO) simulator and matches them with empirical radio signal quality and diverse traffic models. A dataset was created in an urban scenario in the city of Berlin with more than 6h of duration, containing more than 40000 UEs served by 21 cells \cite{oliveira2021generating}.

The development and evaluation of contextual beamforming techniques for vehicle communication in the 5G network rely heavily on datasets. Researchers have utilized various datasets from the literature to evaluate and investigate the effectiveness of these techniques. This systematic review critically analyzes the datasets used by researchers worldwide to incorporate contextual beamforming techniques in vehicular communication. \cite{Phung2022Open} presented the 5G3E dataset, which contains thousands of time series related to the observation of multiple resources involved in 5G network operation. The dataset was created to support 5G network automation. The 5G3E dataset contained thousands of time series related to the observation of multiple resources involved in 5G network operation. The variety of collected features ranged from radio front-end metrics to physical server operating system and network function metrics. The testbed was deployed to support the creation of traffic starting from real traffic traces of a commercial network operator.

Another dataset is the 5G trace dataset from a significant Irish mobile operator introduced by \cite{Raca2020Beyond}. This paper presented a 5G trace dataset collected from a major Irish mobile operator. The dataset was generated from two mobility patterns (static and car) and across two application patterns (video streaming and file download). The dataset was composed of client-side cellular key performance indicators (KPIs) comprised of channel-related metrics, context-related metrics, cell-related metrics and throughput information. Additionally, The authors provided a 5G large-scale multi-cell ns-3 simulation framework to supplement our real-time 5G production network dataset. This framework allowed other researchers to investigate the interaction between users connected to the same cell through the generation of their own synthetic datasets.

\cite{Karim2023SPEC5G} curated SPEC5G, the first publicly accessible 5G dataset for natural language processing (NLP) research. The dataset contains 134M words in 3,547,587 phrases taken from 13 online websites and 13094 cellular network specs. The authors utilized this dataset for security-related text categorization and summarization by utilizing large-scale pre-trained language models. For protocol testing, pertinent security-related attributes were also extracted using text classification techniques. Additionally, \cite{Bouchelaghem2022User} presented a novel mobility dataset generation method for 5G networks based on users' GPS trajectory data. It aggregated the user's GPS trajectories and models his location history by a mobility graph representing the cell base stations he passed through. The generated dataset contained the mobility graph records of 128 users. The user mobility dataset for 5g networks based on GPS geolocation is valuable for predicting user mobility patterns.

In another study, \cite{Lee2022Network} discussed a methodology for collecting a labelled dataset for a 5G network. It described how to build a 5G testbed and use it to collect data. This data can then be used to construct a 5G-based labelled dataset. A 5G testbed was built to observe 5G network features by replaying the collected data. A specialized network collector system was implemented to collect 5G edge network traffic data. A re-collecting methodology using the proposed 5G testbed and network collector can be used to construct a 5G-based labelled dataset for supervised learning methods.

\begin{table}[]
\centering
\label{tab3}
\caption{Table listing datasets related to contextual beamforming}
\resizebox{\columnwidth}{!}{%
\begin{tabular}{|l|l|l|}
\hline
\textbf{Dataset Name}                                                       & \textbf{Source}                                                                    & \textbf{Characteristics}                                                                                                                                         \\ \hline
\begin{tabular}[c]{@{}l@{}}IEEE 802.11n \\ Channel Measurement\end{tabular} & \begin{tabular}[c]{@{}l@{}}IEEE 802.11n \\ working group\end{tabular}              & \begin{tabular}[c]{@{}l@{}}Channel state information (CSI) \\ from multiple antennas\end{tabular}                                                                \\ \hline
KTH Localization Dataset                                                    & \begin{tabular}[c]{@{}l@{}}KTH \\ Royal Institute of Technology\end{tabular}       & \begin{tabular}[c]{@{}l@{}}Received signal strength (RSS) and \\ Angle of arrival (AoA) from multiple antennas\\ Ground truth location data\end{tabular}         \\ \hline
CSI-Hotel Dataset                                                           & \begin{tabular}[c]{@{}l@{}}University of California,\\  Santa Barbara\end{tabular} & \begin{tabular}[c]{@{}l@{}}CSI data from multiple antennas \\ in an indoor environment\end{tabular}                                                              \\ \hline
DeepMIMO Dataset                                                            & New York University                                                                & \begin{tabular}[c]{@{}l@{}}Synthetic data generated by a ray-tracing tool \\ for a variety of scenarios, including urban and \\ indoor environments\end{tabular} \\ \hline
iNEMO Dataset                                                               & STMicroelectronics                                                                 & \begin{tabular}[c]{@{}l@{}}Acceleration, magnetic field, and \\ angular velocity data, \\ along with ground truth location data\end{tabular}                     \\ \hline
\end{tabular}%
}
\end{table}

Also, \cite{Mehmeti2021Analyzing} used the results of a publicly available measurement campaign of 5G users and analyzes various figures of merit. The analysis showed that the downlink and uplink rates for static and mobile users can be captured either by a lognormal or a Generalized Pareto distribution. Downlink and uplink rates for static and mobile users can be captured either by a lognormal or a Generalized Pareto distribution. Time spent in the same cell by a mobile (driving) user can be captured to the best extent by a Generalized Pareto distribution. Prediction of the number of active users in the cell is possible.

Furthermore, \cite{Narayanan20205G} discussed 5G Tracker - a crowdsourced platform that includes an Android app to record passive and active measurements tailored to 5G networks and research. It has been used for over 8 months and has collected over 4 million data points. The platform is useful for building the first-of-a-kind, interactive 5G coverage mapping application. 5G Tracker is a crowdsourced platform to enable research using commercial 5G services. 5G performance is affected by several user-side contextual factors such as user mobility level, orientation, weather, location dynamics, and environmental features. 5G Tracker has been used to collect over 4 million data points, consuming over 50 TB of cellular data across multiple 5G carriers in the U.S.

Moreover, bringing computational and storage technologies closer to end users with strategically deployed and opportunistic processing and storage resources, mobile edge computing in the 5G network was developed by \cite{Lei2019Data} as a very attractive computation architecture. This paper used data mining and statistical methods to analyze Baidu website data. The analysis results gave suggestions to improve the design and development of 5G services. Data mining and statistical analysis of Baidu cloud services in 5G network revealed that clustering, outlier detection, prediction, and statistical methods can be used to evaluate smart city services. The analysis results provided insights into the design and development of 5G services (API website). The findings suggested that mobile edge computing in 5G networks can be used to improve the performance of smart city services.

Finally, \cite{Yang2020Big} discussed a project to collect "three-level" edge layer data from equipment manufacturers, equipment users, and spare parts manufacturers. The goal was to establish a unified data standard and help companies build general services around equipment data. 5G+ Industrial Internet is used to collect data from three levels of edge layers: spare parts manufacturers, equipment manufacturers and equipment users. Data is assimilated and unified into a single standard. A large-scale and shallow informatization project of incremental equipment is implemented in the Yangtze River Delta in a short period of time.
Also, the table lists some 5G data sets consisting of Channel state information (CSI), Phase, Received Power, etc. that can be used for user localisation.

\subsection{Optimization techniques for Contextual Beamforming [RQ:3]}

Contextual beamforming models can be optimised using various strategies to ensure real-time processing. One method to accelerate computations is through hardware acceleration techniques like GPU processing. Additionally, pre-trained models or reducing the number of parameters in the model design can be used to optimise the model. To decrease the model size and boost computational effectiveness, techniques like pruning, quantization, and knowledge distillation can be applied. Improving the feature extraction and input data pre-processing phases can also help with real-time processing. Creating specialised algorithms and optimisation strategies adapted for certain hardware and deployment conditions can also enhance the performance of contextual beamforming models.

Below are some of the possible ways to optimise contextual beamforming models for real-time processing:

\begin{figure}[!t]
 \centering
 \includegraphics[width=8cm]{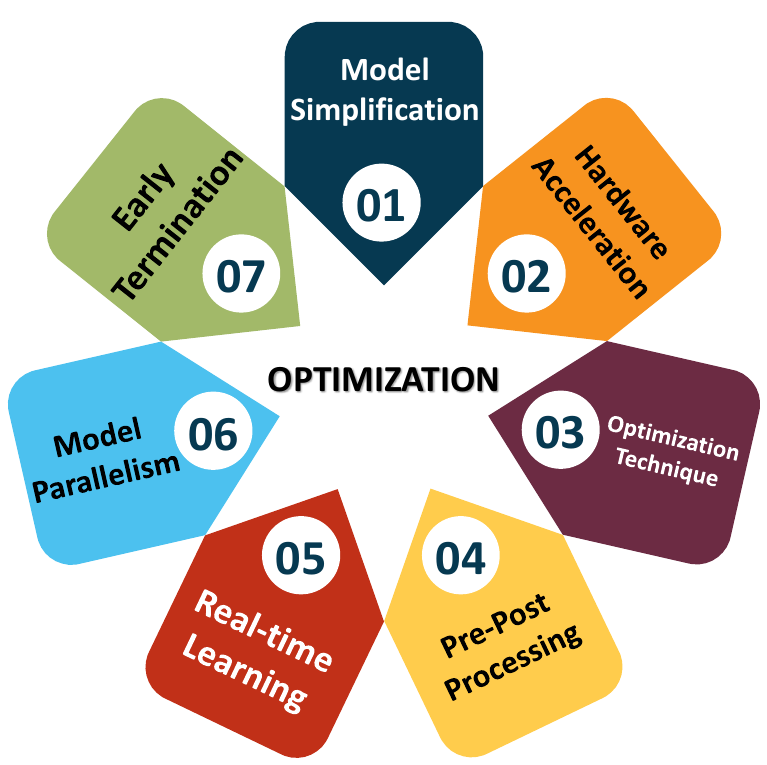}
 \caption{Optimization Techniques for Contextual Beamforming}
 \label{fig:Ref}
\end{figure}

\subsubsection{Model simplification:} Simplifying the model architecture, such as reducing the number of layers or the number of neurons in each layer, can improve computational efficiency and reduce processing time. For instance, the research suggests a strategy that uses machine learning and hardware performance counter data to optimise power and performance for GPU-based systems. The model can accurately detect power-performance bottlenecks and provide optimization techniques for a variety of sophisticated compute and memory access patterns. The model, which has been validated on NVIDIA Fermi C2075 and M2090 GPUs as well as the Keeneland supercomputer at Georgia Tech, is more reliable and accurate than existing GPU power models \cite{song2013simplified}.
    
\subsubsection{Hardware acceleration:} Dedicated hardware, such as graphics processing units (GPUs), field-programmable gate arrays (FPGAs), or application specific integrated circuits (ASICs), can speed up the processing of contextual beamforming models by performing parallel computations. In order to determine the direction of incoming signals, beamforming is a signal processing technique that combines signals from a number of receivers. Although it overcomes noise interference, adaptive beamforming (ABF) is computationally expensive. On current graphics processing units (GPUs), ABF can be implemented in parallel. ABF can be parallelized on an NVIDIA GPU using the author's method, which has a lesser throughput than the serial implementation but is still able to be improved \cite{romer2008beamforming}.
    
\subsubsection{Optimization techniques:} Various optimization techniques, such as weight pruning, quantization, and knowledge distillation, can be applied to contextual beamforming models to reduce their computational complexity and memory footprint without significant loss in accuracy. This research \cite{tang2016recurrent} explores the use of the deep neural network (DNN) model as the teacher to train recurrent neural networks (RNNs), specifically long short-term memory (LSTM), for automated voice recognition (ASR). The method successfully trains RNNs without the use of additional learning methods, even with a small amount of training data.
    
\subsubsection{Preprocessing and postprocessing:} Preprocessing the input data to reduce its dimensionality or complexity, and postprocessing the output data to refine the results or reduce noise, can help improve the performance and efficiency of contextual beamforming models. With the use of deep learning, a low-complexity precoding design approach for multiuser MIMO systems is suggested in \cite{zhang2022deep}. The suggested method uses methods such as input dimensionality reduction, network pruning, and recovery module compression to produce a performance that is comparable to the conventional WMMSE algorithm with relatively little computational cost.
   
\subsubsection{Real-time learning:} Using online or incremental learning algorithms, instead of offline or batch learning, can enable contextual beamforming models to adapt to changing conditions in real-time and reduce the need for frequent retraining. In \cite{wassermann2020adaptive}, two adaptive learning approaches such as ADAM and RAL are proposed for the real-time detection of network assaults in Internet network traffic. These methods achieve excellent detection accuracy even in the presence of idea drifts by dynamically learning from and adapting to non-stationary data streams while lowering the demand for labelled data.
   
\subsubsection{Model parallelism:} Breaking the model into smaller sub-models and processing them in parallel can improve the overall processing speed of contextual beamforming models. This can be done using techniques such as data parallelism or model parallelism. With a focus on model and data parallelization, \cite{verbraeken2020survey} addresses distributed machine learning architecture and topology. It analyses machine learning algorithms and offers parallelization suggestions. The specific needs and demands of communications networks, such as resource allocation and trade-offs between privacy and security, are not addressed.
   
\subsubsection{Early termination:} Stopping the model's processing early when a certain threshold is reached can reduce unnecessary computation, especially in cases where the output has already converged. \cite{prechelt1998automatic} promotes early pausing before convergence to prevent overfitting and suggests the use of cross-validation to detect overfitting during neural network training. The study uses multi-layer perceptrons with RPROP training to assess the effectiveness and efficiency of 14 distinct automatic stopping criteria from three classes for a variety of activities. The findings indicate that slower stopping criteria slightly improve generalisation, although training time often increases by a factor of four.

The choice of optimization techniques will depend on the specific requirements of the application and the constraints of the hardware platform. A combination of these techniques can be used to achieve the best balance between performance and efficiency for the real-time processing of contextual beamforming models.

\subsection{AI, ML and DL approaches for Contextual Beamforming [RQ:4]}

AI, is a term that encompasses a vast array of techniques and technologies that grant machines the ability to perform tasks that would usually demand human intelligence, such as learning, problem-solving, and decision-making. Within AI, there are two primary categories: narrow or weak AI and general or strong AI. Narrow or weak AI machines are designed to accomplish specific tasks, while general or strong AI strives to create machines capable of performing any cognitive task a human can do. In contrast, machine learning (ML) is a subfield of AI that specializes in the development of algorithms and statistical models that enable machines to improve their performance on a task over time by learning from data. ML algorithms can be classified into three primary categories: supervised learning, unsupervised learning, and reinforcement learning. In supervised learning, the algorithm is trained on labelled data, where the correct output is already known, to forecast new outputs for unseen data. In unsupervised learning, the algorithm is trained on unlabeled data to identify patterns or structures in the data. In reinforcement learning, the algorithm learns through trial and error, receiving feedback in the form of rewards or penalties based on its actions.

Regarding beamforming, AI can refer to any technique that allows machines to enhance the quality or efficiency of beamforming by learning from data, making predictions or decisions based on that data, and adapting to changing conditions. Machine learning is a specific subset of AI that utilizes algorithms and statistical models to enable machines to learn from data without explicit programming. The amalgamation of beamforming and artificial intelligence (AI) represents a compelling advancement in signal processing and communication systems. Beamforming, a technique used in radio communications and signal processing, involves the direction of a signal toward a particular location or direction. By integrating signals from multiple antennas, beamforming amplifies signals in the desired direction while suppressing interference from other directions. On the other hand, AI employs computational algorithms and computer programs that can learn from existing data to make decisions or predictions. AI finds application in various domains, including natural language processing (NLP), image recognition, and robotics.

By combining beamforming and AI, communication systems can witness remarkable improvements in their performance. AI algorithms can scrutinize signals received by multiple antennas and determine the optimal beamforming configuration for a given situation. This can result in enhanced signal quality and reduced interference. Additionally, AI can dynamically adjust beamforming parameters in response to current environmental and signal characteristics using reinforcement learning or other AI techniques, which is particularly useful in complex and dynamic environments where traditional beamforming techniques may struggle to adapt. AI is also beneficial for optimizing beamforming algorithms themselves by adjusting the parameters employed to combine signals from different antennas. This can enhance the accuracy and efficiency of the beamforming process, leading to more reliable communication. Furthermore, beamforming and AI can significantly improve the performance of communication systems in various applications, from cellular networks to satellite communication systems.

Recent research has delved into AI-assisted contextual beamforming, which can be optimized using AI algorithms to filter out unwanted noise from the signal or to automatically identify the location of a sound source. This can be accomplished by training models on datasets of sound signals and corresponding locations and using the models to predict the location of new sound sources or to identify and remove noise from new signals. By pointing the microphone array towards the predicted location, sound can be captured more effectively. In multi-user multiple-input-single-output (MISO) systems, beamforming is a useful way to improve the quality of incoming signals. Traditionally, finding the best beamforming solution has relied on iterative techniques, which have significant processing delays and are unsuitable for real-time applications \cite{xia2019deep}. With recent advancements in deep learning (DL) algorithms, identifying the best beamforming (BF) solution in real-time while taking into account both performance and computational delay has become possible. This is accomplished by offline training of neural networks before online optimization, allowing the trained neural network to identify the optimal BF solution. This approach reduces computational complexity during online optimization, requiring only simple linear and nonlinear operations \cite{xia2019deep}.

\begin{figure*}[!t]
    \centering
    \includegraphics[width=\textwidth]{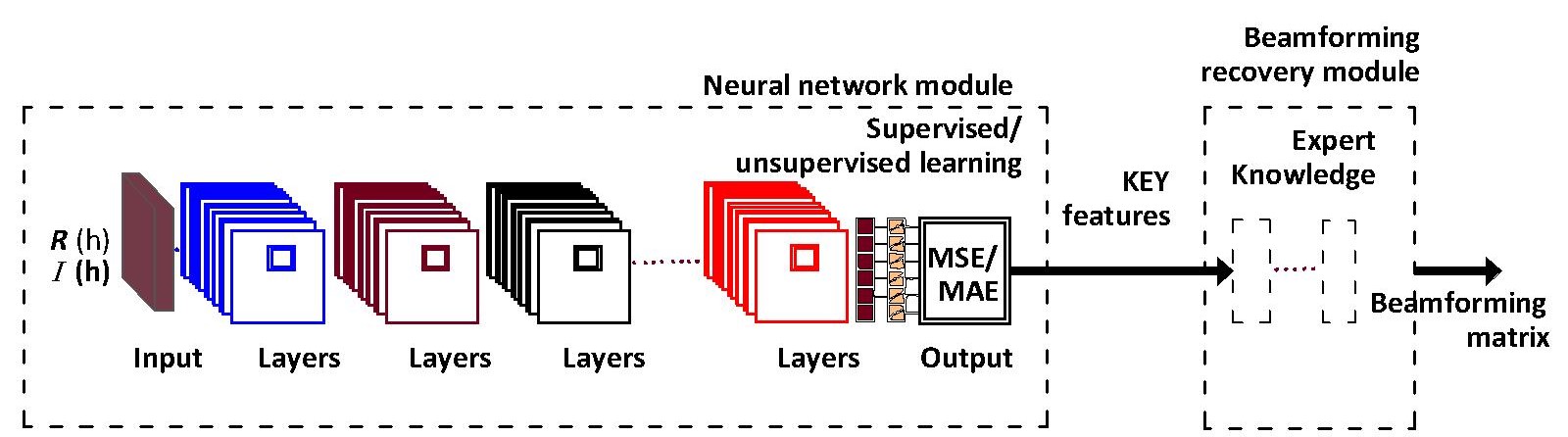}
    \caption{A Basic Architecture of a deep neural network that consists of input (extracted features), neural layers(as per required framework), output (desired results) and a feed for post-processing.}
    \label{neural network}
\end{figure*}

Figure \ref{neural network} illustrates the neural network architecture for BF, which comprises input, neural layers, and output to extract features for further processing. In complicated indoor or outdoor contexts with multiple pathways, propagation loss, noise, and Doppler effects create additional issues. Chong Liu's approach involves employing a machine learning regression method based on efficient BF transmission patterns to predict the position of users on the move, following the collection of large volumes of Line-of-Sight (LOS) and Non-Line-of-Sight (NLoS) data \cite{liu2020improved}. In the domain of location estimation, Bhattacharjee et al. presented two distinct approaches for training neural networks, one using channel parameters as features and the other using a channel response vector, and evaluated the results using preliminary computer simulations \cite{bhattacherjee2020localization, bhattacherjee2021experimental}. The same group also conducted experimental work on the localization of drones and other application areas using different approaches \cite{bhattacherjee2022experimental}.  Wang et al. proposed a weighted loss function to enhance the performance of localization with sparse sensor layouts, achieving an accuracy boost of over 50\% \cite{wang2022deep}. We also presented results for future location estimation of mobile users using a deep neural network in \cite{kaur2022deep}.

Contextual Beamforming in 5G vehicle communication has been implemented using various ML and AI techniques. The performance and precision of beamforming systems, which are essential for efficient communication in moving situations, are to be improved by these techniques. ML has the potential to significantly advance 5G technology, as evidenced by the growing complexity of constructing cellular networks. Deep learning has demonstrated effectiveness in ML tasks like speech recognition and computer vision, with performance growing as more data is accessible. The proliferation of deep learning applications in wireless communications is constrained by the scarcity of huge datasets. To create channel realisations that accurately depict 5G scenarios with mobile transceivers and objects, this study describes an approach that combines a car traffic simulator with a raytracing simulator. The following section of the review offers a unique dataset along with various ML as well as AI techniques used for examining millimetre wave beam selection methods for car-to-infrastructure communication. The application of datasets produced with the suggested methodology is demonstrated by experiments including deep learning in classification, regression, and reinforcement learning problems \cite{Klautau5}.

\begin{table*}[]
\label{tab2}
\caption{Table to compare the performance of different contextual beamforming techniques based on results from various studies}
\resizebox{\textwidth}{!}{%
\begin{tabular}{|l|l|l|l|l|}
\hline
\textbf{Technique Name}                                                          & \textbf{Description}                                                                                                                                  & \textbf{Advantages}                                                                                                              & \textbf{Limitations}                                                                                                                                                                 & \textbf{Type of Data Required}                                                                                       \\ \hline
Geometric-Based                                                                  & \begin{tabular}[c]{@{}l@{}}Determines the location of the user using\\ the arrival times of signals from multiple antennas\end{tabular}               & -Low computational cost                                                                                                          & \begin{tabular}[c]{@{}l@{}}-Limited accuracy in indoor environments\\ -Vulnerable to multipath fading\end{tabular}                                                                   & \begin{tabular}[c]{@{}l@{}}-Antenna array data\\ -User location data\end{tabular}                                    \\ \hline
\begin{tabular}[c]{@{}l@{}}Channel State Information \\ (CSI)-Based\end{tabular} & \begin{tabular}[c]{@{}l@{}}Uses CSI data from multiple antennas \\ to estimate the user's location\end{tabular}                                       & \begin{tabular}[c]{@{}l@{}}-High accuracy\\ -Robustness to multipath fading\end{tabular}                                         & \begin{tabular}[c]{@{}l@{}}-Requires high-quality CSI data\\ -Complex algorithms\end{tabular}                                                                                        & \begin{tabular}[c]{@{}l@{}}-CSI data from multiple antennas\\ -User location data\end{tabular}                       \\ \hline
Hybrid-Based                                                                     & \begin{tabular}[c]{@{}l@{}}Combines geometric and CSI-based \\ techniques to improve accuracy and robustness\end{tabular}                             & \begin{tabular}[c]{@{}l@{}}-High accuracy\\ -Robustness to multipath fading and noise\end{tabular}                               & \begin{tabular}[c]{@{}l@{}}-Requires complex algorithms\\ -May have high computational cost\end{tabular}                                                                             & \begin{tabular}[c]{@{}l@{}}-Antenna array data\\ -CSI data from multiple antennas\\ -User location data\end{tabular} \\ \hline
Machine Learning-Based                                                           & \begin{tabular}[c]{@{}l@{}}Uses machine learning algorithms to \\ learn the relationship between antenna \\ array data and user location\end{tabular} & \begin{tabular}[c]{@{}l@{}}-High accuracy\\ -Can adapt to changing environments\end{tabular}                                     & \begin{tabular}[c]{@{}l@{}}-Requires large amounts of training data\\ -May have high computational cost during training\end{tabular}                                                 & \begin{tabular}[c]{@{}l@{}}-Antenna array data\\ -User location data for training\end{tabular}                       \\ \hline
Deep Learning-Based                                                              & \begin{tabular}[c]{@{}l@{}}Uses deep learning algorithms to learn the \\ relationship between antenna array data and user location\end{tabular}       & \begin{tabular}[c]{@{}l@{}}-High accuracy\\ -Can adapt to changing environments\\ -Can handle large amounts of data\end{tabular} & \begin{tabular}[c]{@{}l@{}}-Requires even larger amounts of training data \\ than machine learning-based techniques\\ -May have high computational cost during training\end{tabular} & \begin{tabular}[c]{@{}l@{}}-Antenna array data\\ -User location data for training\end{tabular}                       \\ \hline
\end{tabular}%
}
\end{table*}

\subsubsection{Deep Learning Techniques:} The extraction of valuable characteristics from input signals and the provision of more precise predictions have been accomplished using deep learning techniques like CNNs and RNNs. 

For instance, Wang et al. used deep learning to simplify beamforming weight estimation in 5G systems. They developed a channel model and trained convolutional neural networks on generated data. The networks predicted beamforming weights based on channel data, reducing complexity. Results show the potential of deep learning for digital and hybrid beamforming, and performance comparison with conventional techniques was presented \cite{aljohani2022implementation}. Also, \cite{silva2022selection} proposed a method that aims to improve the performance of Random Forest, Multilayer Perceptron, and k-Nearest Neighbors classification models by increasing the amount of data through synthetic data inclusion. Their experimental results showed that the inclusion of synthetic data improved the macro F1 scores of the models. The Random Forest, Multilayer Perceptron, and k-Nearest Neighbors achieved macro F1 scores of 0.9341, 0.9241, and 0.9456, respectively, which are higher than those obtained with the original data only.
    
\cite{huang2019fast} proposed a deep learning-based fast-beamforming design method for sum rate maximization under a total power constraint. The method was trained offline using a two-step training strategy. Simulation results demonstrated that the proposed method is fast while obtaining a comparable performance to the state-of-the-art method. They derived a heuristic solution structure of the downlink beamforming through the virtual equivalent uplink channel based on the optimum MMSE receiver. BPNet is designed to perform the joint optimization of power allocation and VUB design and is trained offline using a two-step training strategy. A DL-enabled beamforming neural network (BFNN) is proposed which can optimize the beamformer to attain better spectral efficiency. Simulation findings reveal that the proposed BFNN achieves significant performance gain and high robustness to imperfect CSI. The proposed BFNN greatly decreases the computational complexity compared to conventional BF algorithms. Spectral Efficiency, Performance Gain, Robustness To Imperfect Csi, and Computational Complexity (Measured In Floating Point Operations) are the main outcomes of BFNN \cite{jeyakumar2022beamforming}.
    
\cite{xia2019deep} proposed a beamforming neural network (BNN) for the power minimization problem in multi-antenna communication systems. The BNN was based on convolutional neural networks and the exploitation of expert knowledge. It achieved satisfactory performance with low computational delay. A deep fully convolutional neural network (CNN) was used for beamforming, providing considerable performance gains. The CNN was trained in a supervised manner considering both uplink and downlink transmissions with a loss function based on UE receiver performance. The neural network predicted the channel evolution between uplink and downlink slots and learned to handle inefficiencies and errors in the whole chain, including the actual beamforming phase \cite{huttunen2022deeptx}. A deep learning model that learns how to use these signatures to predict the beamforming vectors at the BSs. \cite{alkhateeb2018deep} discussed a novel integrated machine learning and coordinated beamforming solution to support highly-mobile mmWave applications. The solution used a deep learning model to learn how to use signatures to predict the beamforming vectors at the base stations. This rendered a comprehensive solution that supports highly mobile mmWave applications with reliable coverage, low latency, and negligible training overhead.  
    
\cite{hameed2021deep} proposed a deep learning–based energy beamforming scheme for a multi-antennae wireless powered communication network (WPCN). We used offline training for the deep neural network (DNN) to provide a faster solution to the real-time resource allocation optimization problem. Simulation results showed that the proposed DNN scheme provided a fair approximation of the traditional sequential parametric convex approximation (SPCA) method with low computational and time complexity.

\subsubsection{Supervised ML Techniques:}

Different sorts of acoustic environments have been classified and predicted using supervised learning methods like SVMs and decision trees. A modified SVM technique is proposed for 3D MIMO beamforming in 5G networks. The Advanced Encryption Standard algorithm is employed for more security, and interference is reduced in two stages. The suggested ML-3DIM method outperforms existing methods in terms of throughput, SINR, and SNR by up to 20\%, 30\%, and 35\%, respectively, according to simulation results \cite{yadav20223d}. \cite{kwon2019machine} investigated the machine learning-based beamforming design in two-user MISO interference channels. It proposed a machine learning structure that takes transmit power and channel vectors as input and then recommends two users' choices between MRT and ZF as output. The numerical results showed that our proposed machine learning-based beamforming design well finds the best beamforming combination and achieved a sum rate of more than 99.9\% of the best beamforming combination.
    
\cite{ramon2005beamforming} introduced an SVM-based approach for linear array processing and beamforming. It showed how the new minimization approach can be applied to the problem of linear beamforming. BER performance of LS and SVM for different noise levels from 0 to 15 dB. A machine learning (ML) beamforming approach based on the k-nearest neighbours ( ${k}$ -NN) approximation has been considered, which was trained to generate the appropriate beamforming configurations according to the spatial distribution of throughput demand. Performance was evaluated statistically, via a developed system-level simulator that executes Monte Carlo simulations in parallel. The ML-assisted beamforming framework achieved up to 5 Mbits/J and 36 bps/Hz in terms of energy efficiency (EE) and spectral efficiency (SE), respectively, with reduced hardware and algorithmic complexity \cite{lavdas2022machine}. BeamMaP was a beamforming-based machine learning model for positioning in massive MIMO systems. Simulation results showed that BeamMaP achieved Reduced Root-Mean-Squared Estimation Error (RMSE) performance with an increasing volume of training data. BeamMaP was more efficient and steady in the positioning system compared with kNN and SVM \cite{liu2019beammap}.
    
\cite{singh2020machine} discussed a machine learning method for beamforming at the receiver side antennas for deploying Line-of-Sight (LOS) communication in Satellite Communication (Satcom). It described how the antenna array weights are pre-calculated for a number of beam directions and kept as a database. The signal weights that were calculated for each array element by using their progressive measured phase difference were due to the arriving signal, which was given as input to a linear regression machine learning model and the direction of arrival (DOA) of the signal is predicted. A method for determining an appropriate precoder from the knowledge of the user’s location only was proposed. The proposed method involved a neural network with a specific structure based on random Fourier features allowing to learn functions containing high spatial frequencies. The proposed method was able to handle both line-of-sight (LOS) and non-line-of-sight (NLOS) channels \cite{le2022deep}.

\subsubsection{Unsupervised ML Techniques:} Unsupervised learning methods like clustering and PCA have been used to spot trends and put related data points in one category. For instance, \cite{lin2020unsupervised} proposes a beamforming algorithm for fifth-generation and later communication systems. The approach combines the benefits of conventional optimization-based beamforming techniques with deep learning-based techniques. To create initial beamforming, a novel neural network architecture is proposed, and performance is increased by building a deep unfolding module. The entire algorithm is unsupervised and trained, and simulation results demonstrate enhanced performance and reduced computing complexity when compared to current approaches.
    
\cite{hojatian2022flexible} proposed a novel unsupervised learning approach to design the hybrid beamforming for any subarray structure while supporting quantized phase shifters and noisy CSI. No beamforming codebook was required, and the neural network is trained to take into account the phase-shifter quantization. Simulation results showed that the proposed deep learning solutions can achieve higher sum rates than existing methods.

\subsubsection{Reinforcement ML Techniques:}

Contextual Beamforming system performance has been enhanced using reinforcement learning techniques like Q-learning and policy gradient methods. For instance, to make network design and maintenance more straightforward, a brand-new intelligent algorithm for massive MIMO beamforming performance optimisation is proposed in this research. To produce accurate user mobility patterns, pertinent antenna designs, and an estimate of the effectiveness of the generated antenna diagrams, the system uses three neural networks that apply deep adversarial reinforcement learning workflow. This method has the advantage of learning independently and without requiring big training datasets \cite{maksymyuk2018deep}.
    
\cite{sun2020machine} investigated the use of deep reinforcement learning to predict coordinated beamforming strategy in an ultra-dense network. It was found that the optimal solution is a balanced combination of selfish and altruistic beamforming. The beamforming vectors were obtained efficiently through the learned balancing coefficients. A reinforcement learning (RL) based algorithm for cognitive beamforming was proposed for multi-target detection in massive multiple input multiple outputs (MMIMO) cognitive radars (MMIMO CR). The proposed RL-based algorithm outperformed the conventional omnidirectional approach with equal power allocation in terms of target detection performance. The performance improvement was even more remarkable under environmentally harsh conditions such as low SNR, heavy-tailed disturbance and rapidly changing scenarios \cite{ahmed2020reinforcement}.
    
\cite{raj2022deep} proposed a blind beam alignment method based on RF fingerprints of user equipment obtained from base stations. They used deep reinforcement learning on a multiple-base station cellular environment with multiple mobile users. Achieved a data rate of up to four times the data rate of the traditional method without any overheads. \cite{bai2020multiagent} proposed a novel multiagent reinforcement learning(MARL) formulation for codebook-based beamforming control. It took advantage of the inherently distributed structure in a wirelessly powered network and laid the groundwork for fully locally computed beam control algorithms. A cognitive beamforming algorithm based on Reinforcement Learning (RL) framework is proposed for colocated MIMO radars. The proposed RL-based beamforming algorithm is able to iteratively sense the unknown environment and synthesize a set of transmitted waveforms tailored to the acquired knowledge. The performance of the proposed RL-based beamforming algorithm is assessed in terms of Probability of Detection $(P_{D})$ \cite{wang2018reinforcement}.
    
\cite{nasim2020learning} proposed a reinforcement learning (RL) approach called combinatorial multi-armed bandit (CMAB) framework to maximize the overall network throughput for multi-vehicular communications. They proposed an adaptive combinatorial Thompson sampling algorithm, namely adaptive CTS, and a sequential Thompson sampling (TS) algorithm for the appropriate selection of simultaneous beams in a high-mobility vehicular environment. Simulation results showed that both of our proposed strategies approach the optimal achievable rate achieved by the genie-aided solution.

\begin{table*}[]
\caption{Table summarizing the different types of machine learning and artificial intelligence techniques used in contextual beamforming}
\resizebox{\textwidth}{!}{%
\begin{tabular}{|p{4cm}|l|l|l|}
\hline
\textbf{Technique Name}             & \textbf{Description}                                                                                                                                 & \textbf{Advantages}                                                                                                                                       & \textbf{Limitations}                                                                                                                       \\ \hline
Support Vector Machines (SVM)       & \begin{tabular}[c]{@{}l@{}}Supervised learning algorithm that \\ learns a decision boundary between classes\end{tabular}                             & \begin{tabular}[c]{@{}l@{}}-Can handle high-dimensional data\\ -Effective in binary classification tasks\end{tabular}                                     & -May overfit with noisy or imbalanced data                                                                                                               \\ \hline
Random Forest (RF)                  & \begin{tabular}[c]{@{}l@{}}Ensemble learning method that \\ combines multiple decision trees \\ to improve performance\end{tabular}                  & \begin{tabular}[c]{@{}l@{}}-Can handle high-dimensional data,\\ -Can handle missing or noisy data\\ -Can provide feature importance measures\end{tabular} & -May overfit with noisy or imbalanced data                                                                                                                 \\ \hline
Convolutional Neural Networks (CNN) & \begin{tabular}[c]{@{}l@{}}Neural network architecture that \\ uses convolutional layers to extract \\ features from input data\end{tabular}         & \begin{tabular}[c]{@{}l@{}}-Highly effective for image and signal processing tasks\\ -Can learn complex spatial patterns\end{tabular}                     & \begin{tabular}[c]{@{}l@{}}-May require large amounts of training data\\ -May be computationally expensive\end{tabular}                                  \\ \hline
Recurrent Neural Networks (RNN)     & \begin{tabular}[c]{@{}l@{}}Neural network architecture that \\ can process sequential data by \\ maintaining a memory of past inputs\end{tabular}    & \begin{tabular}[c]{@{}l@{}}-Effective for time-series data and natural language processing tasks\\ -Can handle variable-length inputs\end{tabular}        & \begin{tabular}[c]{@{}l@{}}-May be prone to vanishing or exploding gradients\\ -May require large amounts of training data\end{tabular}                 \\ \hline
Reinforcement Learning (RL)         & \begin{tabular}[c]{@{}l@{}}Learning paradigm in which an \\ agent learns to make decisions \\ through trial and error in an environment\end{tabular} & \begin{tabular}[c]{@{}l@{}}-Can adapt to changing environments\\ -Can handle complex decision-making tasks\end{tabular}                                   & \begin{tabular}[c]{@{}l@{}}-May require significant computational resources\\ -May require careful tuning of hyperparameters\end{tabular}                 \\ \hline
\end{tabular}%
}
\end{table*}

\subsubsection{Hybrid Techniques:}

Contextual Beamforming system performance has also been improved using hybrid methods that incorporate several ML and AI techniques, such as deep reinforcement learning. To address the hybrid beamforming issue in huge MIMO systems, deep reinforcement learning is suggested. The suggested techniques reduce computing complexity while achieving spectral efficiency performance that is close to ideal \cite{arjoune2021double}. Hybrid beamforming, combining digital baseband precoders and analogue RF phase shifters, is an effective technique for millimetre wave (mmWave) communications and massive multiple-in-multiple-out (MIMO) systems. Machine learning techniques can be used to improve the achievable spectral efficiency of hybrid beamforming systems. The proposed two-step algorithm can attain almost the same efficiency as that can be achieved by fully digital architectures \cite{aljumaily2019machine}.
    
\cite{aljumaily2021hybrid} described the design of ML-based hybrid beamforming for multiple users in systems that use millimetre waves (mmWaves) and massive MIMO architectures. The simulation results showed that the ML-based hybrid beamforming architecture can achieve the same spectral efficiency (bits/sec/Hz) as the fully digital beamforming designs with negligible error for both single-user and multi-user Massive-MIMO scenarios. \cite{hojatian2021unsupervised} proposed a novel RSSI-based unsupervised deep learning method to design the hybrid beamforming in massive MIMO systems. They proposed a method to design the synchronization signal (SS) in initial access (IA) and a method to design the codebook for the analog precoder. They showed that the proposed method not only greatly increases the spectral efficiency especially in frequency-division duplex (FDD) communication by using partial CSI feedback, but also has a near-optimal sum-rate and outperforms other state-of-the-art full-CSI solutions.
    
Deep neural networks (DNNs) can be used to approximate the singular value decomposition (SVD) and design hybrid beamformers. DNN-based hybrid beamforming improved rates by up to 50-70\% compared to conventional hybrid beamforming algorithms and achieved a 10-30\% gain in rates compared with the state-of-the-art ML-aided hybrid beamforming algorithms. The proposed approach had low time complexity and memory requirements \cite{peken2020deep}. A federated learning (FL) based framework for hybrid beamforming was proposed, where the model training was performed at the base station (BS) by collecting only the gradients from the users. A convolutional neural network was designed, in which the input was the channel data, yielding the analog beamformers at the output. FL was demonstrated to be more tolerant to the imperfections and corruptions in the channel data as well as having less transmission overhead than centralized machine learning (CML) \cite{elbir2020federated}.

\section{Challenges associated with using AI, ML and DL techniques for contextual beamforming}

Contextual beamforming is a technique used in signal processing and communication systems to improve the quality of sound or data transmission by focusing the transmitted or received signals in a specific direction or area of interest. Machine learning (ML) techniques have been increasingly used to optimize the performance of contextual beamforming systems. However, there are several challenges associated with using ML techniques for contextual beamforming:
\begin{enumerate}
    \item \textbf{Lack of training data}: ML techniques require a large amount of data to be trained effectively. However, in contextual beamforming, it may be difficult to collect enough data that accurately represents the various environments and scenarios in which the system will be used. This can result in underfitting or overfitting of the model, leading to poor performance.
    
    \item \textbf{Complexity of the models}: ML models used for contextual beamforming can be quite complex, with many parameters that need to be tuned. This can make the training process difficult and time-consuming, and can also increase the risk of overfitting.
    
    \item \textbf{Robustness to environmental changes}: Contextual beamforming systems need to be robust to changes in the environment, such as changes in noise levels or the location of sound sources. ML models may not be able to adapt to these changes quickly enough, resulting in reduced performance.
    
    \item \textbf{Limited interpretability}: ML models can be difficult to interpret, which can make it hard to understand why the system is behaving in a certain way or to diagnose problems when they occur.
    
    \item \textbf{Limited generalizability}: ML models trained on one set of data may not generalize well to other datasets or environments. This can limit the applicability of the system in real-world scenarios.

\end{enumerate}

\begin{table*}[]
\caption{Table comparing the computational complexity and processing time of different contextual beamforming models}
\resizebox{\textwidth}{!}{%
\begin{tabular}{|p{2.9cm}|p{2.2cm}|p{1.5cm}|p{5cm}|p{5cm}|p{1.2cm}|}
\hline
\textbf{Model}                                  & \textbf{Computational Complexity} & \textbf{Processing Time} & \textbf{Advantages}                                                                          & \textbf{Limitations}                                                              \\ \hline
Linear Beamforming                              & Low                               & Fast                     & \begin{tabular}[c]{@{}l@{}}-Simple to implement\\ -low computational complexity\end{tabular} & Cannot perform well in non-line-of-sight environments                                             \\ \hline
Maximum Ratio Transmission (MRT)                & Low                               & Fast                     & \begin{tabular}[c]{@{}l@{}}-Simple to implement\\ -low computational complexity\end{tabular} & Cannot perform well in interference-limited environments                                          \\ \hline
Minimum Variance Distortionless Response (MVDR) & High                              & Slow                     & -Provide better performance in non-line-of-sight environments                                & Computationally intensive                                                                            \\ \hline
Neural Network-Based Beamforming                & High                              & Slow                     & -Learn complex non-linear relationships between inputs and outputs                           & Require significant computational resources for training                                              \\ \hline
Reinforcement Learning-Based Beamforming        & High                              & Slow                     & -Adapt to changing environments and optimize performance through trial and error             & Require significant computational resources and careful tuning of hyperparameters              \\ \hline
\end{tabular}%
}
\end{table*}

To address these challenges, researchers are exploring new techniques such as transfer learning, which involves pretraining models on large datasets and then fine-tuning them on smaller, task-specific datasets. They are also working on developing more interpretable ML models and incorporating robustness and adaptability into the models. These challenges have been discussed in several research papers, including Another study proposed by \cite{uddin2018proposing} discussed a novel approach to enhance the Feature Engineering and Selection (eFES) Optimization process in ML. eFES was built using a unique scheme to regulate error bounds and parallelize the addition and removal of a feature during training. Results showed the promising state of eFES as compared to the traditional feature selection process. A weak convolutional network can be used to provide rough label maps over the neighbourhood of a pixel. Incorporating this weak learner in a bigger network can improve the accuracy of state-of-the-art architectures. The approach in \cite{kekecc2014contextually} was generic and can be applied to similar networks where contextual cues are available at training time.

A Multicriteria technique has been developed that allows for the control of feature effects on the model’s output. Knowledge functions have been integrated to accommodate for more complex effects and local lack of information. A Deep Learning training process that was both interpretable and compliant with modern legislation has been developed by \cite{repetto2022multicriteria}. \cite{kim2019structure} proposed a technique to improve the interpretability in transfer learning tasks by defining interpretable features. They examined the interpretability of transfer learning by applying a pre-trained model with defined features to Korean character classification. Feature Network (FN) consisted of Feature Extraction Layer and a single mapping layer that connected the features extracted from the source domain to the target domain. Also, \cite{zhu2017target} proposed an actor-critic model that allowed better generalization across goals and scenes. AI2-THOR framework enabled agents to take actions and interact with objects, allowing for efficient collection of training samples. Model converged faster than state-of-the-art deep reinforcement learning methods, generalized to real robot scenarios with minimal fine-tuning, and is end-to-end trainable.

\section{Potential Applications of Contextual Beamforming}

Some potential applications of contextual beamforming for 5G technology include:

\begin{enumerate}
    \item \textbf{Improved coverage}: Contextual beamforming can help extend the coverage of 5G networks by focusing the transmission beam towards the receiver. This can help overcome obstacles such as buildings and trees that may obstruct the signal.
    
    \item \textbf{Higher data rates}: By directing the signal towards the receiver, contextual beamforming can help increase the data rates of 5G networks. This can enable faster downloads and uploads, as well as smoother streaming of high-definition content.
    
    \item \textbf{Reduced interference}: Contextual beamforming can help reduce interference from other devices or networks by steering the transmission beam away from sources of interference. This can improve the reliability and quality of 5G connections.
   
    \item \textbf{Energy efficiency}: By directing the transmission beam towards the receiver, contextual beamforming can reduce the amount of energy required to transmit the signal. This can help improve the energy efficiency of 5G networks, which is an important consideration for mobile devices that rely on battery power.

\end{enumerate}

\section{Our Contribution Towards Localisation and Beamforming}
 
Optimal BF strikes a balance between giving maximum power to a single user while decreasing or eliminating signal interference at other users. When the maximum ratio transmission (MRT) BF technique is used in an MU-MIMO system, the transmitter transmits a beam to every user according to its weight vector. The resultant power received by each user for the signal intended for that user is calculated as the product of the channel gain and weight vector. Because the MIMO system transmits to multiple users at the same frequency, a critical performance metric for the system is the signal-to-interference-plus-noise ratio (SINR) for each user. This concept has been proved in \cite{kaur2021improving} that shows how SINR can be significantly improved by 28.83 dBm and/ or 53\% by using MRT in comparison with no BF.

We have also published some of the work in context to localization in \cite{kaur2021enhancing, kaur2021improving, kaur2022deep, ahmad2022uav}. These works show how location datasets can be extracted through ray tracing tools and how data can be utilized for location prediction using deep neural networks. In our recent paper(\cite{mallouhi2022network}), the usage of the core network is discussed to enable beamforming after calculating the direction of arrival, or in simple terms, the user's location at the MAC layer. The basic concept of this paper is shown in Figure \ref{crop}.

\section{Conclusion}
In this study, we provide an overview of advanced adaptive BF in which artificial intelligence techniques such as deep learning (DL) can be used. More importantly, we have shown that with access to contextual information such as prior user location, a wireless network's performance can be improved through deep learning techniques. With the development of exciting new technologies such as edge computing and federated learning, we believe the next generation of mobile networks will unlock new opportunities. Communication systems will continue to evolve as closed-loop systems where data extracted by observing a mobile user will be exploited to improve connectivity and network performance, such as the signal-to-noise ratio (SNR). We have touched upon some of the studies already underway that can harness a user's location, and develop a DL-enabled contextual beamforming strategy that can improve the SNR by 53\% on average.

\bibliographystyle{IEEEtran}
\bibliography{ref}

\begin{thebibliography}{100}
\providecommand{\url}[1]{#1}
\csname url@samestyle\endcsname
\providecommand{\newblock}{\relax}
\providecommand{\bibinfo}[2]{#2}
\providecommand{\BIBentrySTDinterwordspacing}{\spaceskip=0pt\relax}
\providecommand{\BIBentryALTinterwordstretchfactor}{4}
\providecommand{\BIBentryALTinterwordspacing}{\spaceskip=\fontdimen2\font plus
\BIBentryALTinterwordstretchfactor\fontdimen3\font minus
  \fontdimen4\font\relax}
\providecommand{\BIBforeignlanguage}[2]{{%
\expandafter\ifx\csname l@#1\endcsname\relax
\typeout{** WARNING: IEEEtran.bst: No hyphenation pattern has been}%
\typeout{** loaded for the language `#1'. Using the pattern for}%
\typeout{** the default language instead.}%
\else
\language=\csname l@#1\endcsname
\fi
#2}}
\providecommand{\BIBdecl}{\relax}
\BIBdecl

\bibitem{chataut2020massive}
R.~Chataut and R.~Akl, ``Massive mimo systems for 5g and beyond
  networks—overview, recent trends, challenges, and future research
  direction,'' \emph{Sensors}, vol.~20, no.~10, p. 2753, 2020.

\bibitem{kumar2021review}
P.~Kumar \emph{et~al.}, ``Review paper on development of mobile wireless
  technology,'' in \emph{Journal of Physics: Conference Series}, vol. 1979,
  no.~1.\hskip 1em plus 0.5em minus 0.4em\relax IOP Publishing, 2021, p.
  012024.

\bibitem{gohar2021role}
A.~Gohar and G.~Nencioni, ``The role of 5g technologies in a smart city: The
  case for intelligent transportation system,'' \emph{Sustainability}, vol.~13,
  no.~9, p. 5188, 2021.

\bibitem{huo20175g}
Y.~Huo, X.~Dong, and W.~Xu, ``5g cellular user equipment: From theory to
  practical hardware design,'' \emph{IEEE Access}, vol.~5, pp.
  13\,992--14\,010, 2017.

\bibitem{zhang2015large}
Z.~Zhang, X.~Wang, K.~Long, A.~V. Vasilakos, and L.~Hanzo, ``Large-scale
  mimo-based wireless backhaul in 5g networks,'' \emph{IEEE Wireless
  Communications}, vol.~22, no.~5, pp. 58--66, 2015.

\bibitem{alhayani20225g}
B.~Alhayani, A.~S. Kwekha-Rashid, H.~B. Mahajan, H.~Ilhan, N.~Uke,
  A.~Alkhayyat, and H.~J. Mohammed, ``5g standards for the industry 4.0 enabled
  communication systems using artificial intelligence: Perspective of smart
  healthcare system,'' \emph{Applied nanoscience}, pp. 1--11, 2022.

\bibitem{sandoval2016evaluating}
R.~M. Sandoval, A.-J. Garcia-Sanchez, F.~Garcia-Sanchez, and J.~Garcia-Haro,
  ``Evaluating the more suitable ism frequency band for iot-based smart grids:
  A quantitative study of 915 mhz vs. 2400 mhz,'' \emph{Sensors}, vol.~17,
  no.~1, p.~76, 2016.

\bibitem{sethi2017internet}
P.~Sethi and S.~R. Sarangi, ``Internet of things: architectures, protocols, and
  applications,'' \emph{Journal of Electrical and Computer Engineering}, vol.
  2017, 2017.

\bibitem{rappaport2013millimeter}
T.~S. Rappaport, S.~Sun, R.~Mayzus, H.~Zhao, Y.~Azar, K.~Wang, G.~N. Wong,
  J.~K. Schulz, M.~Samimi, and F.~Gutierrez, ``Millimeter wave mobile
  communications for 5g cellular: It will work!'' \emph{IEEE access}, vol.~1,
  pp. 335--349, 2013.

\bibitem{attaran2021impact}
M.~Attaran, ``The impact of 5g on the evolution of intelligent automation and
  industry digitization,'' \emph{Journal of Ambient Intelligence and Humanized
  Computing}, pp. 1--17, 2021.

\bibitem{pahlavan2001principles}
K.~Pahlavan and P.~Krishnamurthy, \emph{Principles of wireless networks}.\hskip
  1em plus 0.5em minus 0.4em\relax Prentice Hall PTR, 2001, vol.~1.

\bibitem{akyildiz2018combating}
I.~F. Akyildiz, C.~Han, and S.~Nie, ``Combating the distance problem in the
  millimeter wave and terahertz frequency bands,'' \emph{IEEE Communications
  Magazine}, vol.~56, no.~6, pp. 102--108, 2018.

\bibitem{tripathi2021millimeter}
S.~Tripathi, N.~V. Sabu, A.~K. Gupta, and H.~S. Dhillon, ``Millimeter-wave and
  terahertz spectrum for 6g wireless,'' in \emph{6G Mobile Wireless
  Networks}.\hskip 1em plus 0.5em minus 0.4em\relax Springer, 2021, pp.
  83--121.

\bibitem{marzetta2010noncooperative}
T.~L. Marzetta, ``Noncooperative cellular wireless with unlimited numbers of
  base station antennas,'' \emph{IEEE transactions on wireless communications},
  vol.~9, no.~11, pp. 3590--3600, 2010.

\bibitem{zakrzewska2014towards}
A.~Zakrzewska, S.~Ruepp, and M.~S. Berger, ``Towards converged 5g mobile
  networks-challenges and current trends,'' in \emph{Proceedings of the 2014
  ITU kaleidoscope academic conference: Living in a converged world-Impossible
  without standards?}\hskip 1em plus 0.5em minus 0.4em\relax IEEE, 2014, pp.
  39--45.

\bibitem{sand2009position}
S.~Sand, R.~Tanbourgi, C.~Mensing, and R.~Raulefs, ``Position aware adaptive
  communication systems,'' in \emph{2009 Conference Record of the Forty-Third
  Asilomar Conference on Signals, Systems and Computers}.\hskip 1em plus 0.5em
  minus 0.4em\relax IEEE, 2009, pp. 73--77.

\bibitem{cheng2012location}
M.~Cheng and X.~Fang, ``Location information-assisted opportunistic beamforming
  in lte system for high-speed railway,'' \emph{EURASIP Journal on Wireless
  Communications and Networking}, vol. 2012, no.~1, pp. 1--7, 2012.

\bibitem{ruan2016low}
H.~Ruan and R.~C. de~Lamare, ``Low-complexity robust adaptive beamforming
  algorithms exploiting shrinkage for mismatch estimation,'' \emph{IET Signal
  Processing}, vol.~10, no.~5, pp. 429--438, 2016.

\bibitem{bogale2020adaptive}
T.~E. Bogale, ``Adaptive beamforming and modulation design for 5g v2i
  networks,'' in \emph{2020 10th Annual Computing and Communication Workshop
  and Conference (CCWC)}.\hskip 1em plus 0.5em minus 0.4em\relax IEEE, 2020,
  pp. 0090--0096.

\bibitem{maschietti2017location}
F.~Maschietti, D.~Gesbert, and P.~de~Kerret, ``Location-aided coordinated
  analog precoding for uplink multi-user millimeter wave systems,'' \emph{arXiv
  preprint arXiv:1711.03031}, 2017.

\bibitem{igbafe2019location}
O.~Igbafe, J.~Kang, H.~Wymeersch, and S.~Kim, ``Location-aware beam alignment
  for mmwave communications,'' \emph{arXiv preprint arXiv:1907.02197}, 2019.

\bibitem{myers2020deep}
N.~J. Myers, Y.~Wang, N.~Gonz{\'a}lez-Prelcic, and R.~W. Heath, ``Deep
  learning-based beam alignment in mmwave vehicular networks,'' in \emph{ICASSP
  2020-2020 IEEE International Conference on Acoustics, Speech and Signal
  Processing (ICASSP)}.\hskip 1em plus 0.5em minus 0.4em\relax IEEE, 2020, pp.
  8569--8573.

\bibitem{li2018generative}
X.~Li, A.~Alkhateeb, and C.~Tepedelenlio{\u{g}}lu, ``Generative adversarial
  estimation of channel covariance in vehicular millimeter wave systems,'' in
  \emph{2018 52nd Asilomar Conference on Signals, Systems, and
  Computers}.\hskip 1em plus 0.5em minus 0.4em\relax IEEE, 2018, pp.
  1572--1576.

\bibitem{alkhateeb2018deep}
A.~Alkhateeb, S.~Alex, P.~Varkey, Y.~Li, Q.~Qu, and D.~Tujkovic, ``Deep
  learning coordinated beamforming for highly-mobile millimeter wave systems,''
  \emph{IEEE Access}, vol.~6, pp. 37\,328--37\,348, 2018.

\bibitem{blogh2002third}
J.~S. Blogh, J.~Blogh, and L.~L. Hanzo, \emph{Third-generation systems and
  intelligent wireless networking: smart antennas and adaptive
  modulation}.\hskip 1em plus 0.5em minus 0.4em\relax John Wiley \& Sons, 2002.

\bibitem{monzingo2004introduction}
R.~A. Monzingo and T.~W. Miller, \emph{Introduction to adaptive arrays}.\hskip
  1em plus 0.5em minus 0.4em\relax Scitech publishing, 2004.

\bibitem{li2010mimo}
Q.~Li, G.~Li, W.~Lee, M.-i. Lee, D.~Mazzarese, B.~Clerckx, and Z.~Li, ``Mimo
  techniques in wimax and lte: a feature overview,'' \emph{IEEE Communications
  magazine}, vol.~48, no.~5, pp. 86--92, 2010.

\bibitem{steinhardt1989adaptive}
A.~O. Steinhardt and B.~D. Van~Veen, ``Adaptive beamforming,''
  \emph{International Journal of Adaptive Control and Signal Processing},
  vol.~3, no.~3, pp. 253--281, 1989.

\bibitem{shaikh2015linear}
S.~Shaikh and D.~K. Panda, ``Linear, non-linear adaptive beamforming algorithm
  for smart antenna system,'' in \emph{2015 International Conference on
  Computer, Communication and Control (IC4)}.\hskip 1em plus 0.5em minus
  0.4em\relax IEEE, 2015, pp. 1--4.

\bibitem{reed1974rapid}
I.~S. Reed, J.~D. Mallett, and L.~E. Brennan, ``Rapid convergence rate in
  adaptive arrays,'' \emph{IEEE Transactions on Aerospace and Electronic
  Systems}, no.~6, pp. 853--863, 1974.

\bibitem{vorobyov2014adaptive}
S.~A. Vorobyov, ``Adaptive and robust beamforming,'' in \emph{Academic Press
  Library in Signal Processing}.\hskip 1em plus 0.5em minus 0.4em\relax
  Elsevier, 2014, vol.~3, pp. 503--552.

\bibitem{lukose2010study}
S.~Lukose and M.~Mathurakani, ``A study on various types of beamforming
  algorithms,'' 2010.

\bibitem{werner2002reduced}
S.~Werner \emph{et~al.}, \emph{Reduced complexity adaptive filtering algorithms
  with applications to communications systems}.\hskip 1em plus 0.5em minus
  0.4em\relax Helsinki University of Technology, 2002.

\bibitem{orikumhi2018location}
I.~Orikumhi, J.~Kang, C.~Park, J.~Yang, and S.~Kim, ``Location-aware
  coordinated beam alignment in mmwave communication,'' in \emph{2018 56th
  Annual Allerton Conference on Communication, Control, and Computing
  (Allerton)}.\hskip 1em plus 0.5em minus 0.4em\relax IEEE, 2018, pp. 386--390.

\bibitem{dai2021adaptive}
S.~Dai, ``Adaptive beamforming and switching in smart antenna systems,'' Ph.D.
  dissertation, University of Glasgow, 2021.

\bibitem{parihar2015branch}
R.~Parihar, ``Branch prediction techniques and optimizations,''
  \emph{University of Rochester, NY, USA}, 2015.

\bibitem{kim2018development}
M.~Kim, H.~Momose, and T.~Nakayama, ``Development of link context-aware
  millimeterwave beam switching system using depth-sensor,'' in \emph{2018
  International Symposium on Antennas and Propagation (ISAP)}.\hskip 1em plus
  0.5em minus 0.4em\relax IEEE, 2018, pp. 1--2.

\bibitem{sellami2021outdoor}
A.~Sellami, L.~Nasraoui, and L.~Najjar, ``Outdoor neighbor-assisted
  localization algorithm for massive mimo systems,'' in \emph{2021 IEEE 94th
  Vehicular Technology Conference (VTC2021-Fall)}.\hskip 1em plus 0.5em minus
  0.4em\relax IEEE, 2021, pp. 1--5.

\bibitem{wang2021location}
H.~Wang, Z.~Zhang, H.~Shi, J.~Dang, L.~Wu, W.~Zheng, and L.~Zhao, ``Location
  aware multi-cell mimo communications assisted by optical positioning,'' in
  \emph{2021 13th International Conference on Wireless Communications and
  Signal Processing (WCSP)}.\hskip 1em plus 0.5em minus 0.4em\relax IEEE, 2021,
  pp. 1--5.

\bibitem{dobler2019lama}
H.~D{\"o}bler and B.~Scheuermann, ``Lama: Location-assisted medium access for
  position-beaconing applications,'' in \emph{Proceedings of the 22nd
  International ACM Conference on Modeling, Analysis and Simulation of Wireless
  and Mobile Systems}, 2019, pp. 253--260.

\bibitem{lazarev2019positioning}
V.~Lazarev, G.~Fokin, and I.~Stepanets, ``Positioning for location-aware
  beamforming in 5g ultra-dense networks,'' in \emph{2019 IEEE International
  Conference on Electrical Engineering and Photonics (EExPolytech)}.\hskip 1em
  plus 0.5em minus 0.4em\relax IEEE, 2019, pp. 136--139.

\bibitem{xing2021location}
Z.~Xing, R.~Wang, X.~Yuan, and J.~Wu, ``Location-aware beamforming design for
  reconfigurable intelligent surface aided communication system,'' in
  \emph{2021 IEEE/CIC International Conference on Communications in China
  (ICCC)}.\hskip 1em plus 0.5em minus 0.4em\relax IEEE, 2021, pp. 201--206.

\bibitem{mohammadi2022location}
A.~Mohammadi, M.~Rahmati, and H.~Malik, ``Location-aware beamforming for
  mimo-enabled uav communications: An unknown input observer approach,''
  \emph{IEEE Sensors Journal}, vol.~22, no.~8, pp. 8206--8215, 2022.

\bibitem{zhu2022outage}
B.~Zhu, Z.~Zhang, and J.~Cheng, ``Outage analysis and beamwidth optimization
  for positioning-assisted beamforming,'' \emph{IEEE Communications Letters},
  2022.

\bibitem{liu2020location}
C.~Liu, W.~Yuan, Z.~Wei, X.~Liu, and D.~W.~K. Ng, ``Location-aware predictive
  beamforming for uav communications: A deep learning approach,'' \emph{IEEE
  Wireless Communications Letters}, vol.~10, no.~3, pp. 668--672, 2020.

\bibitem{lu2020positioning}
Y.~Lu, M.~Koivisto, J.~Talvitie, M.~Valkama, and E.~S. Lohan,
  ``Positioning-aided 3d beamforming for enhanced communications in mmwave
  mobile networks,'' \emph{IEEE Access}, vol.~8, pp. 55\,513--55\,525, 2020.

\bibitem{khosravi2022location}
S.~Khosravi, H.~S. Ghadikolaeiy, J.~Zander, and M.~Petrova, ``Location-aided
  beamforming in mobile millimeter-wave networks,'' \emph{arXiv preprint
  arXiv:2205.09887}, 2022.

\bibitem{mesaros2016tut}
A.~Mesaros, T.~Heittola, and T.~Virtanen, ``Tut database for acoustic scene
  classification and sound event detection,'' in \emph{2016 24th European
  Signal Processing Conference (EUSIPCO)}.\hskip 1em plus 0.5em minus
  0.4em\relax IEEE, 2016, pp. 1128--1132.

\bibitem{mesaros2018multi}
------, ``A multi-device dataset for urban acoustic scene classification,''
  \emph{arXiv preprint arXiv:1807.09840}, 2018.

\bibitem{thiemann2013diverse}
J.~Thiemann, N.~Ito, and E.~Vincent, ``The diverse environments multi-channel
  acoustic noise database (demand): A database of multichannel environmental
  noise recordings,'' in \emph{Proceedings of Meetings on Acoustics ICA2013},
  vol.~19, no.~1.\hskip 1em plus 0.5em minus 0.4em\relax Acoustical Society of
  America, 2013, p. 035081.

\bibitem{vincent20164th}
E.~Vincent, S.~Watanabe, J.~Barker, and R.~Marxer, ``The 4th chime speech
  separation and recognition challenge,'' \emph{URL: http://spandh. dcs. shef.
  ac. uk/chime challenge $\{$Last Accessed on 1 August, 2018$\}$}, 2016.

\bibitem{schrank2016deep}
T.~Schrank, L.~Pfeifenberger, M.~Z{\"o}hrer, J.~Stahl, P.~Mowlaee, and
  F.~Pernkopf, ``Deep beamforming and data augmentation for robust speech
  recognition: Results of the 4th chime challenge,'' \emph{Proc. CHiME}, pp.
  18--20, 2016.

\bibitem{christensen2010chime}
H.~Christensen, J.~Barker, N.~Ma, and P.~D. Green, ``The chime corpus: a
  resource and a challenge for computational hearing in multisource
  environments,'' in \emph{Eleventh Annual Conference of the International
  Speech Communication Association}, 2010.

\bibitem{adavanne2019multi}
S.~Adavanne, A.~Politis, and T.~Virtanen, ``A multi-room reverberant dataset
  for sound event localization and detection,'' \emph{arXiv preprint
  arXiv:1905.08546}, 2019.

\bibitem{politis2022starss22}
A.~Politis, K.~Shimada, P.~Sudarsanam, S.~Adavanne, D.~Krause, Y.~Koyama,
  N.~Takahashi, S.~Takahashi, Y.~Mitsufuji, and T.~Virtanen, ``Starss22: A
  dataset of spatial recordings of real scenes with spatiotemporal annotations
  of sound events,'' \emph{arXiv preprint arXiv:2206.01948}, 2022.

\bibitem{khelifi2019named}
H.~Khelifi, S.~Luo, B.~Nour, H.~Moungla, Y.~Faheem, R.~Hussain, and
  A.~Ksentini, ``Named data networking in vehicular ad hoc networks:
  State-of-the-art and challenges,'' \emph{IEEE Communications Surveys \&
  Tutorials}, vol.~22, no.~1, pp. 320--351, 2019.

\bibitem{uppoor2013generation}
S.~Uppoor, O.~Trullols-Cruces, M.~Fiore, and J.~M. Barcelo-Ordinas,
  ``Generation and analysis of a large-scale urban vehicular mobility
  dataset,'' \emph{IEEE Transactions on Mobile Computing}, vol.~13, no.~5, pp.
  1061--1075, 2013.

\bibitem{mesogiti20215g}
I.~Mesogiti, E.~Theodoropoulou, F.~Setaki, G.~Lyberopoulos, A.~Tzanakaki,
  M.~Anastassopoulos, C.~Politi, P.~Papaioannou, C.~Tranoris, S.~Denazis
  \emph{et~al.}, ``5g-victori: Future railway communications requirements
  driving 5g deployments in railways,'' in \emph{Artificial Intelligence
  Applications and Innovations. AIAI 2021 IFIP WG 12.5 International Workshops:
  5G-PINE 2021, AI-BIO 2021, DAAI 2021, DARE 2021, EEAI 2021, and MHDW 2021,
  Hersonissos, Crete, Greece, June 25--27, 2021, Proceedings}.\hskip 1em plus
  0.5em minus 0.4em\relax Springer, 2021, pp. 21--30.

\bibitem{bassbouss20215g}
L.~Bassbouss, M.~B. Fadhel, S.~Pham, A.~Chen, S.~Steglich, E.~Troudt,
  M.~Emmelmann, J.~Guti{\'e}rrez, N.~Maletic, E.~Grass \emph{et~al.},
  ``5g-victori: Optimizing media streaming in mobile environments using mmwave,
  nbmp and 5g edge computing,'' in \emph{Artificial Intelligence Applications
  and Innovations. AIAI 2021 IFIP WG 12.5 International Workshops: 5G-PINE
  2021, AI-BIO 2021, DAAI 2021, DARE 2021, EEAI 2021, and MHDW 2021,
  Hersonissos, Crete, Greece, June 25--27, 2021, Proceedings}.\hskip 1em plus
  0.5em minus 0.4em\relax Springer, 2021, pp. 31--38.

\bibitem{coronado20195g}
E.~Coronado, S.~N. Khan, and R.~Riggio, ``5g-empower: A software-defined
  networking platform for 5g radio access networks,'' \emph{IEEE Transactions
  on Network and Service Management}, vol.~16, no.~2, pp. 715--728, 2019.

\bibitem{riley2010ns}
G.~F. Riley and T.~R. Henderson, ``The ns-3 network simulator,'' \emph{Modeling
  and tools for network simulation}, pp. 15--34, 2010.

\bibitem{perrone2013design}
L.~F. Perrone, T.~R. Henderson, M.~J. Watrous, and V.~D. Felizardo, ``The
  design of an output data collection framework for ns-3,'' in \emph{2013
  Winter Simulations Conference (WSC)}.\hskip 1em plus 0.5em minus 0.4em\relax
  IEEE, 2013, pp. 2984--2995.

\bibitem{sperling2019mobility}
J.~Sperling, S.~E. Young, V.~Garikapati, A.~L. Duvall, and J.~Beck, ``Mobility
  data and models informing smart cities,'' National Renewable Energy
  Lab.(NREL), Golden, CO (United States), Tech. Rep., 2019.

\bibitem{tosi2017cell}
D.~Tosi, ``Cell phone big data to compute mobility scenarios for future smart
  cities,'' \emph{International Journal of Data Science and Analytics}, vol.~4,
  pp. 265--284, 2017.

\bibitem{alkhateeb2022deepsense}
A.~Alkhateeb, G.~Charan, T.~Osman, A.~Hredzak, J.~Morais, U.~Demirhan, and
  N.~Srinivas, ``Deepsense 6g: A large-scale real-world multi-modal sensing and
  communication dataset,'' \emph{arXiv preprint arXiv:2211.09769}, 2022.

\bibitem{uvaydov2021deepsense}
D.~Uvaydov, S.~D’Oro, F.~Restuccia, and T.~Melodia, ``Deepsense: Fast
  wideband spectrum sensing through real-time in-the-loop deep learning,'' in
  \emph{IEEE INFOCOM 2021-IEEE Conference on Computer Communications}.\hskip
  1em plus 0.5em minus 0.4em\relax IEEE, 2021, pp. 1--10.

\bibitem{oliveira2021generating}
A.~Oliveira and T.~Vaz{\~a}o, ``Generating synthetic datasets for mobile
  wireless networks with sumo,'' in \emph{Proceedings of the 19th ACM
  international symposium on mobility management and wireless access}, 2021,
  pp. 33--42.

\bibitem{Phung2022Open}
\BIBentryALTinterwordspacing
C.-D. Phung, N.-E.-H. Yellas, S.~B. Ruba, and S.~Secci, ``An {Open} {Dataset}
  for {Beyond}-5g {Data}-driven {Network} {Automation} {Experiments},'' in
  \emph{2022 1st {International} {Conference} on 6G {Networking}
  (6GNet)}.\hskip 1em plus 0.5em minus 0.4em\relax IEEE, jul 6 2022. [Online].
  Available: \url{http://dx.doi.org/10.1109/6gnet54646.2022.9830292}
\BIBentrySTDinterwordspacing

\bibitem{Raca2020Beyond}
\BIBentryALTinterwordspacing
D.~Raca, D.~Leahy, C.~J. Sreenan, and J.~J. Quinlan, ``Beyond throughput, the
  next generation,'' in \emph{Proceedings of the 11th {ACM} {Multimedia}
  {Systems} {Conference}}.\hskip 1em plus 0.5em minus 0.4em\relax ACM, may 27
  2020. [Online]. Available: \url{http://dx.doi.org/10.1145/3339825.3394938}
\BIBentrySTDinterwordspacing

\bibitem{Karim2023SPEC5G}
\BIBentryALTinterwordspacing
I.~Karim, K.~S. Mubasshir, M.~M. Rahman, and E.~Bertino, ``Spec5g: A {Dataset}
  for 5g {Cellular} {Network} {Protocol} {Analysis},'' \emph{Archived}, 2023.
  [Online]. Available: \url{https://arxiv.org/abs/2301.09201}
\BIBentrySTDinterwordspacing

\bibitem{Bouchelaghem2022User}
\BIBentryALTinterwordspacing
S.~Bouchelaghem, H.~Boudjelaba, M.~Omar, and M.~Amad, ``User {Mobility}
  {Dataset} for 5g {Networks} {Based} on {GPS} {Geolocation},'' in \emph{2022
  {IEEE} 27th {International} {Workshop} on {Computer} {Aided} {Modeling} and
  {Design} of {Communication} {Links} and {Networks} ({CAMAD})}.\hskip 1em plus
  0.5em minus 0.4em\relax IEEE, nov 2 2022. [Online]. Available:
  \url{http://dx.doi.org/10.1109/CAMAD55695.2022.9966906}
\BIBentrySTDinterwordspacing

\bibitem{Lee2022Network}
\BIBentryALTinterwordspacing
G.~Lee, J.~Lee, Y.~Kim, and J.-G. Park, ``Network {Flow} {Data} {Re}-collecting
  {Approach} {Using} 5g {Testbed} for {Labeled} {Dataset},'' in \emph{2022 24th
  {International} {Conference} on {Advanced} {Communication} {Technology}
  ({ICACT})}.\hskip 1em plus 0.5em minus 0.4em\relax IEEE, feb 13 2022.
  [Online]. Available: \url{http://dx.doi.org/10.23919/ICACT53585.2022.9728877}
\BIBentrySTDinterwordspacing

\bibitem{Mehmeti2021Analyzing}
\BIBentryALTinterwordspacing
F.~Mehmeti and T.~F.~L. Porta, ``Analyzing a 5g {Dataset} and {Modeling}
  {Metrics} of {Interest},'' in \emph{2021 17th {International} {Conference} on
  {Mobility}, {Sensing} and {Networking} ({MSN})}.\hskip 1em plus 0.5em minus
  0.4em\relax IEEE, 12 2021. [Online]. Available:
  \url{http://dx.doi.org/10.1109/MSN53354.2021.00027}
\BIBentrySTDinterwordspacing

\bibitem{Narayanan20205G}
\BIBentryALTinterwordspacing
A.~Narayanan, E.~Ramadan, J.~Quant, P.~Ji, F.~Qian, and Z.-L. Zhang, ``5g
  tracker,'' in \emph{Proceedings of the {SIGCOMM} '20 {Poster} and {Demo}
  {Sessions}}.\hskip 1em plus 0.5em minus 0.4em\relax ACM, aug 10 2020.
  [Online]. Available: \url{http://dx.doi.org/10.1145/3405837.3411394}
\BIBentrySTDinterwordspacing

\bibitem{Lei2019Data}
\BIBentryALTinterwordspacing
Y.~Lei, L.~Qiang, Z.~Yonghao, L.~Hao, Y.~Peng, F.~Lei, L.~Wenjing, and
  Q.~Xuesong, ``Data {Mining} and {Statistical} {Analysis} on {Smart} {City}
  {Services} {Based} on 5g {Network},'' in \emph{2019 15th {International}
  {Wireless} {Communications} amp; {Mobile} {Computing} {Conference}
  ({IWCMC})}.\hskip 1em plus 0.5em minus 0.4em\relax IEEE, 6 2019. [Online].
  Available: \url{http://dx.doi.org/10.1109/IWCMC.2019.8766454}
\BIBentrySTDinterwordspacing

\bibitem{Yang2020Big}
\BIBentryALTinterwordspacing
Q.~Yang, C.~zhang, C.~Hu, H.~Zhou, and H.~Lou, ``Big data collection and
  application based on 5g {Industrial} {Internet} {Three}-level edge layer,''
  \emph{Journal of Physics: Conference Series}, vol. 1684, no.~1, p. 012025,
  nov 1 2020. [Online]. Available:
  \url{http://dx.doi.org/10.1088/1742-6596/1684/1/012025}
\BIBentrySTDinterwordspacing

\bibitem{song2013simplified}
S.~Song, C.~Su, B.~Rountree, and K.~W. Cameron, ``A simplified and accurate
  model of power-performance efficiency on emergent gpu architectures,'' in
  \emph{2013 IEEE 27th International Symposium on Parallel and Distributed
  Processing}.\hskip 1em plus 0.5em minus 0.4em\relax IEEE, 2013, pp. 673--686.

\bibitem{romer2008beamforming}
M.~Romer, ``Beamforming adaptive arrays with graphics processing units,'' Ph.D.
  dissertation, The University of Texas at Austin, 2008.

\bibitem{tang2016recurrent}
Z.~Tang, D.~Wang, and Z.~Zhang, ``Recurrent neural network training with dark
  knowledge transfer,'' in \emph{2016 IEEE international conference on
  acoustics, speech and signal processing (ICASSP)}.\hskip 1em plus 0.5em minus
  0.4em\relax IEEE, 2016, pp. 5900--5904.

\bibitem{zhang2022deep}
M.~Zhang, J.~Gao, and C.~Zhong, ``A deep learning-based framework for low
  complexity multiuser mimo precoding design,'' \emph{IEEE Transactions on
  Wireless Communications}, vol.~21, no.~12, pp. 11\,193--11\,206, 2022.

\bibitem{wassermann2020adaptive}
S.~Wassermann, T.~Cuvelier, P.~Mulinka, and P.~Casas, ``Adaptive and
  reinforcement learning approaches for online network monitoring and
  analysis,'' \emph{IEEE Transactions on Network and Service Management},
  vol.~18, no.~2, pp. 1832--1849, 2020.

\bibitem{verbraeken2020survey}
J.~Verbraeken, M.~Wolting, J.~Katzy, J.~Kloppenburg, T.~Verbelen, and J.~S.
  Rellermeyer, ``A survey on distributed machine learning,'' \emph{Acm
  computing surveys (csur)}, vol.~53, no.~2, pp. 1--33, 2020.

\bibitem{prechelt1998automatic}
L.~Prechelt, ``Automatic early stopping using cross validation: quantifying the
  criteria,'' \emph{Neural networks}, vol.~11, no.~4, pp. 761--767, 1998.

\bibitem{xia2019deep}
W.~Xia, G.~Zheng, Y.~Zhu, J.~Zhang, J.~Wang, and A.~P. Petropulu, ``A deep
  learning framework for optimization of miso downlink beamforming,''
  \emph{IEEE Transactions on Communications}, vol.~68, no.~3, pp. 1866--1880,
  2019.

\bibitem{liu2020improved}
C.~Liu and H.~J. Helgert, ``An improved adaptive beamforming-based machine
  learning method for positioning in massive mimo systems,''
  \emph{International Journal On Advances in Internet Technology}, vol.~6, no.
  1-2, pp. 1--12, 2020.

\bibitem{bhattacherjee2020localization}
U.~Bhattacherjee, C.~K. Anjinappa, L.~Smith, E.~Ozturk, and I.~Guvenc,
  ``Localization with deep neural networks using mmwave ray tracing
  simulations,'' in \emph{2020 SoutheastCon}.\hskip 1em plus 0.5em minus
  0.4em\relax IEEE, 2020, pp. 1--8.

\bibitem{bhattacherjee2021experimental}
U.~Bhattacherjee, \emph{Experimental Study of Localizing an Aerial User and
  Application of Deep Neural Network in Localization}.\hskip 1em plus 0.5em
  minus 0.4em\relax North Carolina State University, 2021.

\bibitem{bhattacherjee2022experimental}
U.~Bhattacherjee, E.~Ozturk, O.~Ozdemir, I.~Guvenc, M.~L. Sichitiu, and H.~Dai,
  ``Experimental study of outdoor uav localization and tracking using passive
  rf sensing,'' in \emph{Proceedings of the 15th ACM Workshop on Wireless
  Network Testbeds, Experimental evaluation \& CHaracterization}, 2022, pp.
  31--38.

\bibitem{wang2022deep}
W.~Wang, B.~Li, Z.~Huang, and L.~Zhu, ``Deep learning-based localization with
  urban electromagnetic and geographic information,'' \emph{Wireless
  Communications and Mobile Computing}, vol. 2022, 2022.

\bibitem{kaur2022deep}
J.~Kaur, O.~R. Popoola, M.~A. Imran, Q.~H. Abbasi, and H.~T. Abbas, ``Deep
  neural network for localization of mobile users using raytracing,''
  \emph{Archived}, 2022.

\bibitem{Klautau5}
{A. Klautau} and {Pedro Batista}, ``5 {G} {MIMO} {Data} for {Machine}
  {Learning} : Application to {Beam}-{Selection} using {Deep} {Learning}.''

\bibitem{aljohani2022implementation}
K.~Aljohani, I.~Elshafiey, and A.~Al-Sanie, ``Implementation of deep learning
  in beamforming for 5g mimo systems,'' in \emph{2022 39th National Radio
  Science Conference (NRSC)}, vol.~1.\hskip 1em plus 0.5em minus 0.4em\relax
  IEEE, 2022, pp. 188--195.

\bibitem{silva2022selection}
D.~H. Silva, D.~A. Ribeiro, M.~A. Ram{\'\i}rez, R.~L. Rosa, S.~Chaudhary, and
  D.~Z. Rodr{\'\i}guez, ``Selection of beamforming in 5g mimo scenarios using
  machine learning approach,'' in \emph{2022 19th International Conference on
  Electrical Engineering/Electronics, Computer, Telecommunications and
  Information Technology (ECTI-CON)}.\hskip 1em plus 0.5em minus 0.4em\relax
  IEEE, 2022, pp. 1--4.

\bibitem{huang2019fast}
H.~Huang, Y.~Peng, J.~Yang, W.~Xia, and G.~Gui, ``Fast beamforming design via
  deep learning,'' \emph{IEEE Transactions on Vehicular Technology}, vol.~69,
  no.~1, pp. 1065--1069, 2019.

\bibitem{jeyakumar2022beamforming}
P.~Jeyakumar, N.~Tharanitaran, E.~Malar, and P.~Muthuchidambaranathan,
  ``Beamforming design with fully connected analog beamformer using deep
  learning,'' \emph{INTERNATIONAL JOURNAL OF COMMUNICATION SYSTEMS}, vol.~35,
  no.~7, 2022.

\bibitem{huttunen2022deeptx}
J.~M. Huttunen, D.~Korpi \emph{et~al.}, ``Deeptx: Deep learning beamforming
  with channel prediction,'' \emph{arXiv preprint arXiv:2202.07998}, 2022.

\bibitem{hameed2021deep}
I.~Hameed, P.~V. Tuan, and I.~Koo, ``Deep learning--based energy beamforming
  with transmit power control in wireless powered communication networks,''
  \emph{IEEE Access}, vol.~9, pp. 142\,795--142\,803, 2021.

\bibitem{yadav20223d}
R.~Yadav and A.~Tripathi, ``3d mimo beamforming using spatial distance svm
  algorithm and interference mitigation for 5g wireless communication
  network,'' \emph{Journal of Cases on Information Technology (JCIT)}, vol.~24,
  no.~4, pp. 1--26, 2022.

\bibitem{kwon2019machine}
H.~J. Kwon, J.~H. Lee, and W.~Choi, ``Machine learning-based beamforming in
  two-user miso interference channels,'' in \emph{2019 International Conference
  on Artificial Intelligence in Information and Communication (ICAIIC)}.\hskip
  1em plus 0.5em minus 0.4em\relax IEEE, 2019, pp. 496--499.

\bibitem{ramon2005beamforming}
M.~M. Ram{\'o}n, N.~Xu, and C.~G. Christodoulou, ``Beamforming using support
  vector machines,'' \emph{IEEE Antennas and Wireless Propagation Letters},
  vol.~4, pp. 439--442, 2005.

\bibitem{lavdas2022machine}
S.~Lavdas, P.~K. Gkonis, Z.~Zinonos, P.~Trakadas, L.~Sarakis, and
  K.~Papadopoulos, ``A machine learning adaptive beamforming framework for 5g
  millimeter wave massive mimo multicellular networks,'' \emph{IEEE Access},
  vol.~10, pp. 91\,597--91\,609, 2022.

\bibitem{liu2019beammap}
C.~Liu and H.~J. Helgert, ``Beammap: Beamforming-based machine learning for
  positioning in massive mimo systems,'' in \emph{Proceedings of the Eleventh
  International Conference on Evolving Internet, Rome, Italy}, vol.~30, 2019.

\bibitem{singh2020machine}
A.~J. Singh and M.~Jayakumar, ``Machine learning based digital beamforming for
  line-of-sight optimization in satcom on the move technology,'' in \emph{2020
  4th International Conference on Electronics, Communication and Aerospace
  Technology (ICECA)}.\hskip 1em plus 0.5em minus 0.4em\relax IEEE, 2020, pp.
  422--427.

\bibitem{le2022deep}
L.~Le~Magoarou, T.~Yassine, S.~Paquelet, and M.~Crussi{\`e}re, ``Deep learning
  for location based beamforming with nlos channels,'' in \emph{ICASSP
  2022-2022 IEEE International Conference on Acoustics, Speech and Signal
  Processing (ICASSP)}.\hskip 1em plus 0.5em minus 0.4em\relax IEEE, 2022, pp.
  8812--8816.

\bibitem{lin2020unsupervised}
C.-H. Lin, Y.-T. Lee, W.-H. Chung, S.-C. Lin, and T.-S. Lee, ``Unsupervised
  resnet-inspired beamforming design using deep unfolding technique,'' in
  \emph{GLOBECOM 2020-2020 IEEE Global Communications Conference}.\hskip 1em
  plus 0.5em minus 0.4em\relax IEEE, 2020, pp. 1--7.

\bibitem{hojatian2022flexible}
H.~Hojatian, J.~Nadal, J.-F. Frigon, and F.~Leduc-Primeau, ``Flexible
  unsupervised learning for massive mimo subarray hybrid beamforming,'' in
  \emph{GLOBECOM 2022-2022 IEEE Global Communications Conference}.\hskip 1em
  plus 0.5em minus 0.4em\relax IEEE, 2022, pp. 3833--3838.

\bibitem{maksymyuk2018deep}
T.~Maksymyuk, J.~Gazda, O.~Yaremko, and D.~Nevinskiy, ``Deep learning based
  massive mimo beamforming for 5g mobile network,'' in \emph{2018 IEEE 4th
  International Symposium on Wireless Systems within the International
  Conferences on Intelligent Data Acquisition and Advanced Computing Systems
  (IDAACS-SWS)}.\hskip 1em plus 0.5em minus 0.4em\relax IEEE, 2018, pp.
  241--244.

\bibitem{sun2020machine}
C.~Sun, Z.~Shi, and F.~Jiang, ``A machine learning approach for beamforming in
  ultra dense network considering selfish and altruistic strategy,'' \emph{IEEE
  Access}, vol.~8, pp. 6304--6315, 2020.

\bibitem{ahmed2020reinforcement}
A.~M. Ahmed, A.~A. Ahmad, S.~Fortunati, A.~Sezgin, M.~S. Greco, and F.~Gini,
  ``Reinforcement learning based beamforming for massive mimo radar
  multi-target detection,'' \emph{arXiv preprint arXiv:2005.04708}, 2020.

\bibitem{raj2022deep}
V.~Raj, N.~Nayak, and S.~Kalyani, ``Deep reinforcement learning based blind
  mmwave mimo beam alignment,'' \emph{IEEE Transactions on Wireless
  Communications}, vol.~21, no.~10, pp. 8772--8785, 2022.

\bibitem{bai2020multiagent}
L.~Bai and Z.~Pang, ``Multiagent reinforcement learning based energy
  beamforming control,'' \emph{arXiv preprint arXiv:2006.08829}, 2020.

\bibitem{wang2018reinforcement}
L.~Wang, S.~Fortunati, M.~S. Greco, and F.~Gini, ``Reinforcement learning-based
  waveform optimization for mimo multi-target detection,'' in \emph{2018 52nd
  Asilomar Conference on Signals, Systems, and Computers}.\hskip 1em plus 0.5em
  minus 0.4em\relax IEEE, 2018, pp. 1329--1333.

\bibitem{nasim2020learning}
I.~Nasim, A.~S. Ibrahim, and S.~Kim, ``Learning-based beamforming for
  multi-user vehicular communications: a combinatorial multi-armed bandit
  approach,'' \emph{IEEE Access}, vol.~8, pp. 219\,891--219\,902, 2020.

\bibitem{arjoune2021double}
Y.~Arjoune and S.~Faruque, ``Double deep q-learning and sac based hybrid
  beamforming for 5g and beyond millimeter-wave systems,'' in \emph{2021 IEEE
  International Conference on Electro Information Technology (EIT)}.\hskip 1em
  plus 0.5em minus 0.4em\relax IEEE, 2021, pp. 422--428.

\bibitem{aljumaily2019machine}
M.~S. Aljumaily and H.~Li, ``Machine learning aided hybrid beamforming in
  massive-mimo millimeter wave systems,'' in \emph{2019 IEEE International
  Symposium on Dynamic Spectrum Access Networks (DySPAN)}.\hskip 1em plus 0.5em
  minus 0.4em\relax IEEE, 2019, pp. 1--6.

\bibitem{aljumaily2021hybrid}
------, ``Hybrid beamforming for multiuser mimo mm wave systems using
  artificial neural networks,'' in \emph{2021 International Conference on
  Advanced Computer Applications (ACA)}.\hskip 1em plus 0.5em minus 0.4em\relax
  IEEE, 2021, pp. 150--155.

\bibitem{hojatian2021unsupervised}
H.~Hojatian, J.~Nadal, J.-F. Frigon, and F.~Leduc-Primeau, ``Unsupervised deep
  learning for massive mimo hybrid beamforming,'' \emph{IEEE Transactions on
  Wireless Communications}, vol.~20, no.~11, pp. 7086--7099, 2021.

\bibitem{peken2020deep}
T.~Peken, S.~Adiga, R.~Tandon, and T.~Bose, ``Deep learning for svd and hybrid
  beamforming,'' \emph{IEEE Transactions on Wireless Communications}, vol.~19,
  no.~10, pp. 6621--6642, 2020.

\bibitem{elbir2020federated}
A.~M. Elbir and S.~Coleri, ``Federated learning for hybrid beamforming in
  mm-wave massive mimo,'' \emph{IEEE Communications Letters}, vol.~24, no.~12,
  pp. 2795--2799, 2020.

\bibitem{uddin2018proposing}
M.~F. Uddin, J.~Lee, S.~Rizvi, and S.~Hamada, ``Proposing enhanced feature
  engineering and a selection model for machine learning processes,''
  \emph{Applied Sciences}, vol.~8, no.~4, p. 646, 2018.

\bibitem{kekecc2014contextually}
T.~Keke{\c{c}}, R.~Emonet, E.~Fromont, A.~Tr{\'e}meau, and C.~Wolf,
  ``Contextually constrained deep networks for scene labeling,'' in
  \emph{British Machine Vision Conference, 2014}, 2014, pp. to--appear.

\bibitem{repetto2022multicriteria}
M.~Repetto, ``Multicriteria interpretability driven deep learning,''
  \emph{Annals of Operations Research}, pp. 1--15, 2022.

\bibitem{kim2019structure}
D.~Kim, W.~Lim, M.~Hong, and H.~Kim, ``The structure of deep neural network for
  interpretable transfer learning,'' in \emph{2019 IEEE International
  Conference on Big Data and Smart Computing (BigComp)}.\hskip 1em plus 0.5em
  minus 0.4em\relax IEEE, 2019, pp. 1--4.

\bibitem{zhu2017target}
Y.~Zhu, R.~Mottaghi, E.~Kolve, J.~J. Lim, A.~Gupta, L.~Fei-Fei, and A.~Farhadi,
  ``Target-driven visual navigation in indoor scenes using deep reinforcement
  learning,'' in \emph{2017 IEEE international conference on robotics and
  automation (ICRA)}.\hskip 1em plus 0.5em minus 0.4em\relax IEEE, 2017, pp.
  3357--3364.

\bibitem{kaur2021improving}
J.~Kaur, O.~R. Popoola, M.~A. Imran, Q.~H. Abbasi, and H.~T. Abbas, ``Improving
  throughput for mobile receivers using adaptive beamforming,'' in \emph{2021
  1st International Conference on Microwave, Antennas \& Circuits
  (ICMAC)}.\hskip 1em plus 0.5em minus 0.4em\relax IEEE, 2021, pp. 1--4.

\bibitem{kaur2021enhancing}
J.~Kaur, Q.~H. Abbasi, A.~B. Sharif, O.~Popoola, M.~A. Imran, and H.~T. Abbas,
  ``Enhancing wave propagation via contextual beamforming,'' in \emph{2021 IEEE
  International Symposium on Antennas and Propagation and USNC-URSI Radio
  Science Meeting (APS/URSI)}.\hskip 1em plus 0.5em minus 0.4em\relax IEEE,
  2021, pp. 781--782.

\bibitem{ahmad2022uav}
I.~Ahmad, J.~Kaur, H.~T. Abbas, Q.~H. Abbasi, A.~Zoha, M.~A. Imran, and
  S.~Hussain, ``Uav-assisted 5g networks for optimised coverage under dynamic
  traffic load,'' in \emph{2022 IEEE International Symposium on Antennas and
  Propagation and USNC-URSI Radio Science Meeting (AP-S/URSI)}.\hskip 1em plus
  0.5em minus 0.4em\relax IEEE, 2022, pp. 1692--1693.

\bibitem{mallouhi2022network}
H.~Mallouhi, J.~Kaur, H.~T. Abbas, and S.~Laki, ``In-network angle
  approximation for supporting adaptive beamforming,'' in \emph{Proceedings of
  the 5th International Workshop on P4 in Europe}, 2022, pp. 61--66.

\end{thebibliography}

\end{document}